\documentclass[cernpreprint,USenglish]{na61doc}
\usepackage{amsmath}
\usepackage{url}

\usepackage{colortbl}
\definecolor{darkred}{rgb}{0.5,0,0}
\definecolor{darkblue}{rgb}{0,0,0.5}
\definecolor{firebrick}{rgb}{0.75,0.125,0.125}
\definecolor{darkgreen}{rgb}{0,0.5,0}

\usepackage[colorlinks=true,linkcolor=firebrick,citecolor=darkgreen,urlcolor=darkblue]{hyperref}

\usepackage{xspace}
\usepackage{enumerate}

\ShineTitle{Two-particle correlations in azimuthal angle and pseudorapidity in inelastic p+p interactions at the CERN Super Proton Synchrotron}

\PreprintIdNumber{CERN-EP-2016-234}

\ShineJournal{Eur. Phys. J. C}	

\ShineJournalRef{Published in: Eur. Phys. J. C77, 59 (2017)}
\ShineDOI{10.1140/epjc/s10052-017-4599-x}

\ShineAbstract{
  Results on two-particle $\Delta\eta\Delta\phi$ correlations in inelastic p+p interactions at 20, 31, 40, 80, and 158~\GeVc are presented. The measurements were performed using the large acceptance \NASixtyOne hadron spectrometer at the CERN Super Proton Synchrotron. The data show structures which can be attributed mainly to effects of resonance decays, momentum conservation, and quantum statistics. The results are compared with the \Epos and UrQMD models.
}

\newcommand{\eV}{\ensuremath{\mbox{e\kern-0.1em V}}\xspace}
\newcommand{\GeV}{\ensuremath{\mbox{Ge\kern-0.1em V}}\xspace}
\newcommand{\MeV}{\ensuremath{\mbox{Me\kern-0.1em V}}\xspace}
\newcommand{\GeVc}{\ensuremath{\mbox{Ge\kern-0.1em V}\!/\!c}\xspace}
\newcommand{\GeVcc}{\ensuremath{\mbox{Ge\kern-0.1em V}\!/\!c^2}\xspace}
\newcommand{\AGeV}{\ensuremath{A\,\mbox{Ge\kern-0.1em V}}\xspace}
\newcommand{\AGeVc}{\ensuremath{A\,\mbox{Ge\kern-0.1em V}\!/\!c}\xspace}
\newcommand{\MeVc}{\ensuremath{\mbox{Me\kern-0.1em V}/c}\xspace}

\newcommand{\dd}{\ensuremath{{\text{d}}}\xspace}
\newcommand{\dedx}{\ensuremath{\dd E\!/\!\dd x}\xspace}

\newcommand{\Geant}{{\scshape Geant}\xspace}

\newcommand{\Epos}{{\scshape Epos}\xspace}

\newcommand{\CernVM}{\textsc{Cern\-\kern-0.05emVM}\xspace}

\begin{document}

\maketitle

\section{Introduction and motivation}

This paper presents experimental results on two-particle correlations in pseudorapidity and azimuthal angle of charged particles produced in inelastic p+p interactions at 20, 31, 40, 80, and 158~\GeVc. The measurements were performed by the multi-purpose NA61/SHINE~\cite{shine,Abgrall:2014fa} experiment at the CERN Super Proton Synchrotron (SPS). They are part of the strong interactions programme devoted to the study of the properties of the onset of deconfinement and search for the critical point of strongly interacting matter. Within this program a two-dimensional scan in collision energy and size of colliding nuclei is in progress. Data on p+p, Be+Be, and Ar+Sc collisions were already recorded and data on p+Pb and Xe+La collisions will be registered within the coming years. The expected signal of a critical point is a non-monotonic dependence of various fluctuation measures in such a scan; for a recent review see Ref.~\cite{Gazdzicki:2015ska}.

Apart from looking for QGP signatures, it is of interest to study specific physical phenomena that happen during and after the collision. The two-particle correlation analysis in pseudorapidity ($\eta$) and azimuthal angle ($\phi$) allows to disentangle different correlation sources which can be directly connected with phenomena like jets, collective flow, resonance decays, quantum statistics effects, conservation laws, etc.

Measurements of two-particle correlations in pseudorapidity and azimuthal angle were first published by the ACM collaboration at the Intersecting Storage Rings (ISR)~\cite{Eggert:1974ek}. Two- and three-body decays of resonances ($\eta$, $\rho^0$, $\omega$) were found to provide the dominant contributions. Two structures were discovered: an enhancement near $\Delta\phi = \pi$ (away-side) explained by the two-body decay scenario and another enhancement at $\Delta\phi \approx 0$ together with an azimuthal ridge (centered at $\Delta\eta \approx 0$) consistent with three-body decays.\footnote{$\Delta\eta$ and $\Delta\phi$ definitions are in Eq.~\ref{eq:deta_dphi}.} These features were confirmed at the higher collision energies of RHIC by the PHOBOS~\cite{Alver:2007wy} collaboration. 

At RHIC and the LHC parton scattering processes become important. In addition to high transverse momentum jets, studies of $\Delta\eta\Delta\phi$ correlations in p+p interactions as well as in collisions of heavy nuclei ~\cite{Porter:2005rc, Porter:2004jt, Abelev:2014mva, Alver:2008aa} found prominent structures explained as arising from the production of minijets, producing a large correlation peak at small opening angles $(\Delta\eta,\Delta\phi) \approx (0,0)$ and a broad structure along $\Delta\eta$ at $\Delta\phi \approx \pi$ (also referred to as away-side ridge). At SPS energies, however, contributions from hard and semi-hard scattering processes are expected to be much smaller.

Collective flow effects also provide significant contributions to two-particle correlations. In non-central nucleus--nucleus collisions they appear as a modulation in the distribution of $\Delta\phi$ both at SPS~\cite{Appelshauser:1997dg} and higher collider energies~\cite{Ackermann:2000tr}. Recently a similar behaviour was observed in high-multiplicity p+Pb collisions~\cite{CMS:2012qk} and even p+p reactions~\cite{Khachatryan:2015lva} at the LHC. The origin of the effect in p+p reactions is not yet understood.
At SPS energies sufficiently high particle multiplicities cannot be produced in p+p collisions. Thus effects of collective flow will not be discussed further.

In this paper the pseudorapidity variable $\eta$ is calculated as $\eta = -\ln(\tan(\Theta/2))$, where $\tan(\Theta) = p_T/p_L$ with $p_T$ the transverse ($x,y$) and
$p_L$ the longitudinal ($z$) component of the particle momentum in the collision centre-of-mass system. In Lorentz transformation of $p_L$ (from NA61/SHINE originally used laboratory system to centre-of-mass system) the pion mass was assumed for all particles.
The azimuthal angle $\phi$ is the angle between the transverse momentum vector and the horizontal ($x$) axis.

The paper is organized as follows. The correlation function is defined in Sec.~\ref{sec:detadphi}.
The experimental setup is presented in Sec.~\ref{sec:setup}.  Data processing and simulation
are described in Sec.~\ref{sec:data_proc}. Data selection and analysis are discussed in Sec.~\ref{sec:datasets}.
Results are presented in Sec.~\ref{sec:results} and compared to
model calculations in Sec.~\ref{sec:models}. A summary in Sec.~\ref{sec:summary} closes the paper.


\section{Two-particle correlations in pseudorapidity and azimuthal angle}\label{sec:detadphi}

Correlations studied in this paper were calculated as a function of the difference in pseudorapidity
($\eta$) and azimuthal angle ($\phi$) between two particles produced in the same event:
\begin{equation}
  \label{eq:deta_dphi}
  \Delta\eta = |{\eta}_1 - {\eta}_2|,
    \hspace{2cm}
    \Delta\phi = |{\phi}_1 - {\phi}_2|.
\end{equation}

The correlation function $C(\Delta\eta,\Delta\phi)$ is defined and calculated as:
\begin{equation}
  \label{eq:correlations}
  C(\Delta\eta,\Delta\phi)=
  \frac{N_{\text{mixed}}^{\text{pairs}}}{N_{\text{data}}^{\text{pairs}}}
  \frac{D(\Delta\eta,\Delta\phi)}{M(\Delta\eta,\Delta\phi)},
\end{equation}
where
  \begin{equation*}
  D(\Delta\eta,\Delta\phi)=\frac{d^2N_{\text{data}}}{d \Delta \eta d
    \Delta \phi}, \hspace{0.5cm} 
  M(\Delta\eta,\Delta\phi)=\frac{d^2N_{\text{mixed}}}{d \Delta \eta d \Delta
    \phi}
\label{eq:D_and_M}
\end{equation*}
are the distributions of particle pairs from data and mixed events (representing uncorrelated background), respectively. The mixed events were constructed using the mixing procedure described in Sec.~\ref{sec:selection}. Distributions $D(\Delta\eta,\Delta\phi)$ and $M(\Delta\eta,\Delta\phi)$ were obtained by accumulating the number of pairs in intervals of $\Delta\eta$ and $\Delta\phi$. For the calculation of $C(\Delta\eta,\Delta\phi)$ both distributions were normalised to the number of pairs ($N^{\text{pairs}}_{\text{data}}$, $N^{\text{pairs}}_{\text{mixed}}$) in the given distribution.
The measurements are restricted to $0 \leq \Delta\eta \leq 3$ and $0 \leq \Delta\phi < \pi$. Finally the correlation functions were mirrored around the point $(\Delta\eta,\Delta\phi) = (0,0)$ and plotted in the range $ -\frac{\pi}{2} < \Delta\phi < \frac{3 \pi}{2}$.

In this paper the correlation function $C(\Delta\eta,\Delta\phi)$ was calculated for charged primary hadrons produced in strong and electromagnetic processes with transverse momentum $p_T < 1.5$~\GeVc in the NA61/SHINE acceptance (for details see~\cite{ppm_edms}).

While CMS~\cite{Khachatryan:2010gv}, ATLAS~\cite{ATLAS:2012ap} and ALICE~\cite{Janik:2012ya} used equivalent definitions, other quantities have also been used to measure two-particle correlations. The ACM collaboration~\cite{Eggert:1974ek} at the ISR studied
\begin{center}
$C^{\text{II}}(\Delta\eta,\Delta\phi)=\langle (n-1)(\rho_n^{\text{II}}(\eta_1,\phi_1,\eta_2,\phi_2) - \rho_n^{\text{I}}(\eta_1,\phi_1)\rho_n^{\text{I}}(\eta_2,\phi_2) \rangle$,
\end{center}
where n is the multiplicity in the event, $\rho_n^{\text{II}}$ the pair density, $\rho_n^{\text{I}}$ the inclusive density and the averaging runs over the events.
 
PHOBOS~\cite{Alver:2007wy} employed the observable 
\begin{center}
$R(\Delta\eta,\Delta\phi)=\langle (n-1)(\rho_n^{\text{II}}(\Delta\eta,\Delta\phi) / \rho_n^{\text{mixed}}(\Delta\eta,\Delta\phi) - 1) \rangle$
\end{center} 
with $\rho_n^{\text{II}}$ and $\rho_n^{\text{mixed}}$ the pair density in data and mixed events, respectively, both normalised to unity.

The STAR collaboration~\cite{Porter:2005rc} used the quantity
\begin{center}
$\Delta\rho / \sqrt{\rho_{\text{mixed}}} = 
(\rho(\Delta\eta,\Delta\phi) - \rho_{\text{mixed}}(\Delta\eta,\Delta\phi) ) / \sqrt{\rho_{\text{mixed}}(\Delta\eta,\Delta\phi)}$,
\end{center}
where $\rho$ and $\rho_{\text{mixed}}$ are the normalised densities of data and mixed pairs respectively.

        
\section{Experimental setup, beams, target, triggers}\label{sec:setup}

\begin{figure*}
  \centering
  \includegraphics[width=0.8\textwidth]{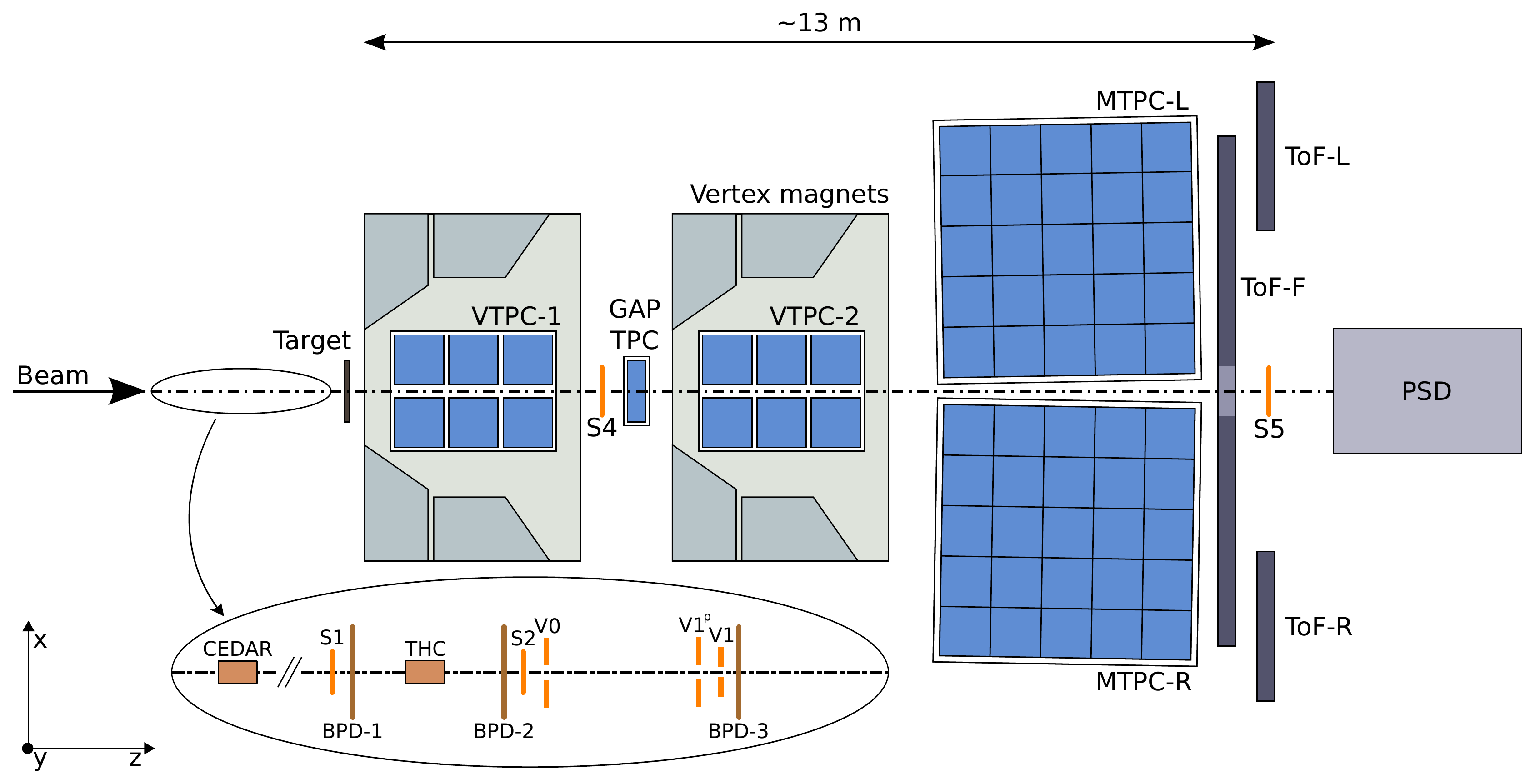}
  \caption[]{
    (Color online) The schematic layout of the NA61/SHINE experiment at the CERN SPS (horizontal cut, not to scale).
  }
  \label{fig:detector-setup}
\end{figure*}

The \NASixtyOne experiment uses a large acceptance hadron spectrometer located in the CERN North Area~\cite{Abgrall:2014fa}. The schematic layout of the \NASixtyOne detector is shown in Fig.~\ref{fig:detector-setup}.

A set of scintillation and Cherenkov counters as well as beam position detectors (BPDs) upstream of the spectrometer provide timing reference, identification and position measurements of incoming beam particles. The trigger scintillator counter S4 placed downstream of the target is used to select events with collisions in the target area by the absence of a charged particle hit. 

Secondary beams of positively charged hadrons at 20, 31, 40, 80, and 158~\GeVc are produced from 400~\GeVc protons extracted from the SPS accelerator. Particles of the secondary hadron beam are identified by two Cherenkov counters, a CEDAR \cite{CEDAR} (either CEDAR-W or CEDAR-N) and a threshold counter (THC). The CEDAR counter, using a coincidence of six out of the eight photo-multipliers placed radially along the Cherenkov ring, provides positive identification of protons, while the THC, operated at pressure lower than the proton threshold, is used in anti-coincidence in the trigger logic. Due to their limited range of operation two different CEDAR counters were used, namely for beams at  20, 31, and 40~\GeVc the CEDAR-W counter and for beams at 80 and 158~\GeVc the CEDAR-N counter. The threshold counter was used for the lower three beam momenta. A selection based on signals from the Cherenkov counters allowed one to identify beam protons with a purity of about 99\%. A consistent value for the purity was found by bending the beam into the TPCs with the full magnetic field and using identification based on its specific ionization energy loss \dedx~\cite{Claudia}.

The main tracking devices of the spectrometer are four large volume Time Projection Chambers (TPCs). Two of them, the vertex TPCs (\mbox{VTPC-1} and \mbox{VTPC-2}), are located in the magnetic fields of two super-conducting dipole magnets with a maximum combined bending power of 9~Tm which corresponds to about 1.5~T and 1.1~T fields in the upstream and downstream magnets, respectively. In order to optimize the acceptance of the detector, the field in both magnets was adjusted proportionally to the beam momentum.

Two large \textit{main} TPCs (\mbox{MTPC-L} and \mbox{MTPC-R}) are positioned downstream of the magnets
symmetrically to the beam line. The fifth small TPC (GAP-TPC) is placed between \mbox{VTPC-1} and \mbox{VTPC-2} directly on the beam line. It closes the gap between the beam axis and the sensitive volumes of the other TPCs. The TPCs are filled with Ar and CO$_2$ gas mixtures. Particle identification in the TPCs is based on measurements of the specific energy loss (\dedx) in the chamber gas and is augmented by mass measurements using Time-of-Flight (ToF) detectors. 

The p+p data, which is the topic of this paper, was taken by colliding the proton beam with a liquid hydrogen target (LHT), a 20~cm long cylinder positioned about 80~cm upstream of \mbox{VTPC-1}.


\section{Data processing, simulation and detector performance}
\label{sec:data_proc}

Detector parameters were optimised by a data-based calibration procedure which also took into account their time dependence. Small adjustments were determined in consecutive steps for:
\begin{enumerate}[(i)]
  \item detector geometry, TPC drift velocities and distortions due to the magnetic field inhomogeneities in the corners of the VTPCs,
  \item magnetic field setting,
  \item specific energy loss measurements.
\end{enumerate}
Each step involved reconstruction of the data required to optimise a given set of calibration constants and time dependent corrections followed by verification procedures. Details of the procedure and quality assessment are presented in Ref.~\cite{Abgrall:1955138}. The resulting performance in the measurements of quantities relevant for this paper is discussed below.

The main steps of the data reconstruction procedure are:
\begin{enumerate}[(i)]
  \item cluster finding in the TPC raw data, calculation of the cluster centre-of-gravity and total charge,
  \item reconstruction of local track segments in each TPC separately,
  \item matching of track segments into global tracks,
  \item track fitting through the magnetic field and determination of track parameters at the first measured TPC cluster,
  \item determination of the interaction vertex using the beam trajectory ($x$ and $y$ coordinates) fitted in the BPDs and the trajectories of tracks reconstructed in the TPCs  ($z$ coordinate),
  \item refitting the particle trajectory using the interaction vertex as an additional point and determining the particle momentum at the interaction vertex.
\end{enumerate}

A simulation of the \NASixtyOne detector response is used to correct the reconstructed data. For this purpose inelastic p+p collisions generated with the \Epos1.99~\cite{Werner:2005jf} model were used to obtain the corrections for reconstruction inefficiency and trigger bias of the \NASixtyOne detector and obtain final corrected results.
\\
The simulation consists of the following steps (see Ref.~\cite{Abgrall:2011rma} for more details):

\begin{enumerate}[(i)]
  \item generation of inelastic p+p interactions using the \Epos model,
  \item propagation of outgoing particles through the detector material using the GEANT 3.21 package~\cite{Geant3} which takes into account the magnetic field as well as relevant physics processes, such as particle interactions and decays,
  \item simulation of the detector response using dedicated \NASixtyOne packages which introduce distortions corresponding to all corrections applied to the real data,
  \item simulation of the interaction trigger selection by checking whether a charged particle hits the S4 counter, see Sec.~\ref{sec:setup},
  \item storage of the simulated events in a file which has the same format as the raw data,
  \item reconstruction of the simulated events with the same reconstruction chain as used for the real data and
  \item matching of reconstructed with simulated tracks based on the cluster positions.
\end{enumerate}

It should be underlined that only inelastic p+p interactions in the hydrogen within the target cell were simulated and reconstructed. Thus the simulation-based corrections can be applied only for inelastic events. The contribution of elastic events is removed by the event selection cuts (see Sec.~\ref{sec:selection}).


\section{Data selection and analysis}\label{sec:datasets}

This section presents the procedures used for data analysis~\cite{BMthesis}. These consist of the following steps: applying event and particle selections, obtaining uncorrected experimental results, evaluation of correction factors based on simulations, and finally calculation of the corrected results.

These steps are described in the subsections below.

\subsection{Event and particle selection}\label{sec:selection}

\subsubsection{Event selection criteria}
\label{sec:event_cuts}

The events selected for the analysis had to satisfy the conditions:

\begin{enumerate}[(i)]
\item event is recorded with the interaction trigger,
\item the beam particle trajectory is measured in at least one of BPD-1 or BPD-2 and in the BPD-3 detector,
\item no off-time beam particle is detected within $\pm 1$~$\mu$s around the trigger particle,
\item there is at least one track reconstructed in the TPCs and fitted to the interaction vertex,
\item the vertex $z$ position (fitted using the beam and TPC tracks) is not farther away than 10~cm from the centre of the LHT,
\item events with a positively charged track with absolute momentum ($p$) close to the beam momentum ($p > p_{\text{beam}} - 3$~\GeVc) are rejected to exclude elastic interactions.
\end{enumerate}

The data statistics for analysed reactions is shown in Table~\ref{tab:data_event_cuts}.

\subsubsection{Track selection criteria and acceptance}
\label{sec:track_cuts}

The tracks selected for the analysis had to satisfy the conditions:

\begin{enumerate}[(i)]
\item track fit converged,
\item the total number of reconstructed points on the track should be at least 30,
\item the sum of the number of reconstructed points in VTPC-1 and VTPC-2 
  should be at least 15 or the number of reconstructed points in the GAP-TPC should be at least five,
\item the distance between the track extrapolated to the interaction plane and the interaction point (impact parameter) should be smaller than 4~cm in the horizontal (bending) plane and 2~cm in the vertical (drift) plane,
\item transverse momentum of particle track should be lower than 1.5~\GeVc,
\item tracks with \dedx and total momentum values characteristic for electrons are rejected (see Ref.~\cite{Abgrall:2013qoa} for the details of this cut).
\end{enumerate}

Model simulations were performed in $4\pi$ acceptance, thus the \NASixtyOne detector acceptance filter needs to be applied before comparisons with data. The detector acceptance was defined as a three-dimensional matrix ($p$,$p_T$,$\phi$) of 1 and 0 depending on whether the bin was or was not populated by particles reconstructed and accepted in the events (see Ref.~\cite{ppm_edms}).

\begin{table}
  \centering
  \begin{tabular}{|l|c|c|c|}
    \hline
    & No cuts &  Events with int. trigger & Cuts applied \\
    \hline
    20 \GeVc & 1320 k & 1094 k & 176 k (13\%) \\
    31 \GeVc & 3134 k & 2828 k & 756 k (24\%) \\
    40 \GeVc & 5238 k & 4683 k & 1444 k (28\%) \\
    80 \GeVc & 4500 k & 3774 k & 1343 k (30\%) \\
    158 \GeVc & 3537 k & 2856 k & 1373 k (39\%) \\
    \hline
  \end{tabular}  
  \caption{Event statistics before and after event cuts for p+p data. See text for detailed description of the cuts.}
  \label{tab:data_event_cuts}
\end{table}

\subsection{Raw correlation function}\label{sec:correlations}

Uncorrected two-particle correlation functions were evaluated using Eq.~\ref{eq:correlations} for events and tracks selected according to the criteria stated above. This was done by calculating differences between each accepted track in pseudorapidity and azimuthal angle with every other track in the same event. The differences were evaluated in the centre-of-mass (CMS) frame and accumulated in two-dimensional bins of ($\Delta\eta$, $\Delta\phi$). To increase statistics, the $\Delta\phi$ range was folded, i.e. for $\Delta\phi$ larger than $\pi$ its value was recalculated as $2\pi - \Delta\phi$. This is allowed due to the symmetry of the p+p reactions. The transformation from laboratory (LAB) to CMS frame was performed assuming the pion mass for all produced particles.

\begin{figure*}
  \centering
  \includegraphics[width=0.3\textwidth]{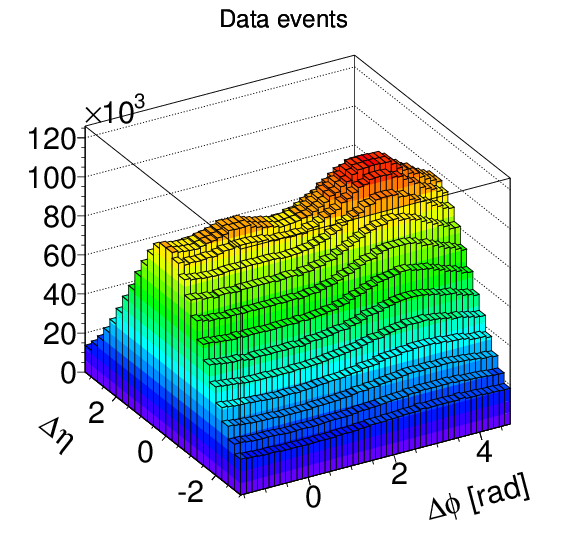}
  \includegraphics[width=0.3\textwidth]{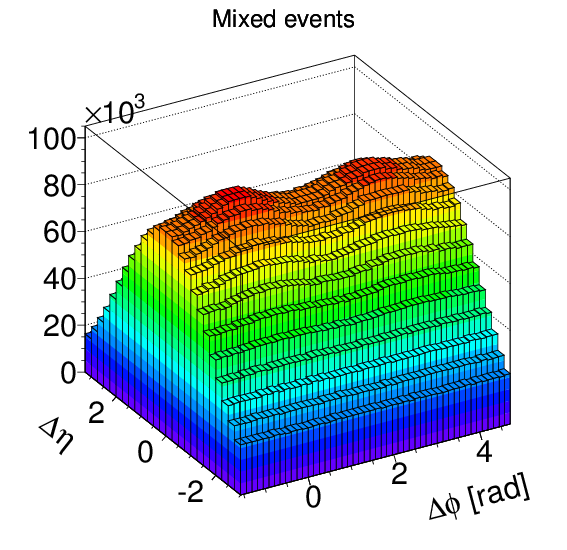}
  \caption{Example distributions of number of pairs of all charged particles in ($\Delta\eta$, $\Delta\phi$) from the data (left) and mixed (right) events in p+p interactions in 158~\GeVc.}
  \label{fig:before_after_mixing}
\end{figure*}

The uncorrelated reference was constructed by mixing particles between events with two main constraints: (a) the multiplicity distribution of mixed data had to be exactly the same as the original; (b) mixed events could not contain two particles from the same original (data) event. Example distributions of number of pairs of charged particles in bins of ($\Delta\eta$, $\Delta\phi$) before and after mixing are shown in Fig.~\ref{fig:before_after_mixing}. The overall shape of the pair distribution in Fig.~\ref{fig:before_after_mixing} (left) results from the shape of the inclusive single particle distribution and the hump structure is caused by the acceptance limitations of the detector. As shown in Fig.~\ref{fig:before_after_mixing} (right) these features are reproduced in the mixed pair distribution, which does not contain effects of true correlations by construction. Thus in the correlation function defined as the ratio in Eq.~\ref{eq:correlations} these background effects cancel and only the true correlation effects remain.

\begin{figure*}[ht]
  \centering
  \includegraphics[width=0.25\textwidth]{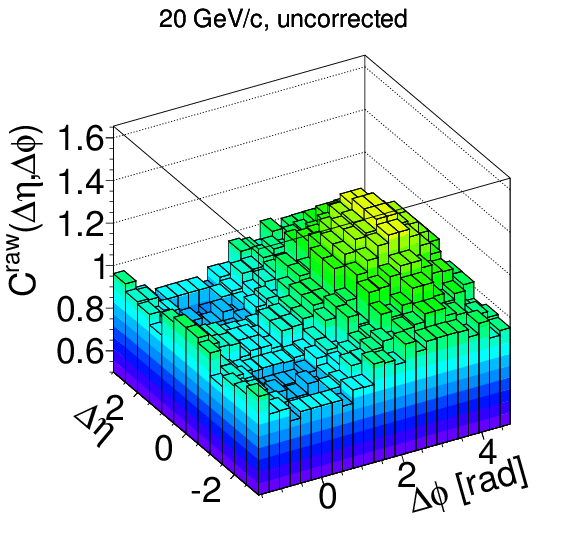}
  \includegraphics[width=0.25\textwidth]{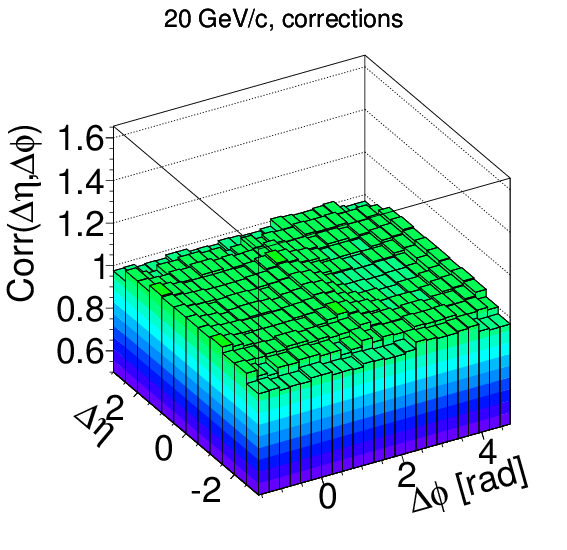}
  \includegraphics[width=0.25\textwidth]{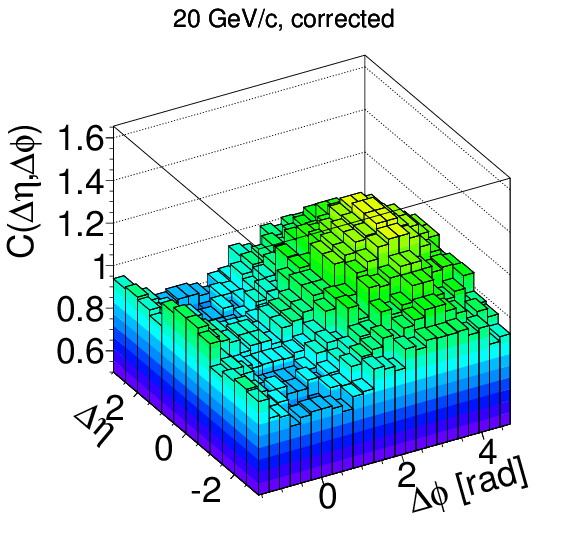}
  \\
  \includegraphics[width=0.25\textwidth]{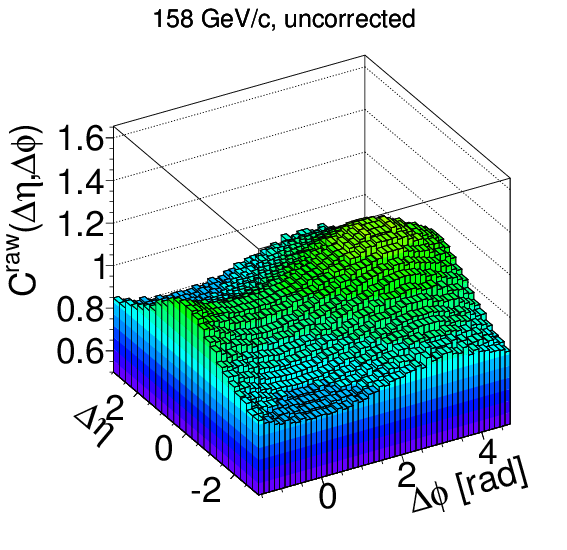}
  \includegraphics[width=0.25\textwidth]{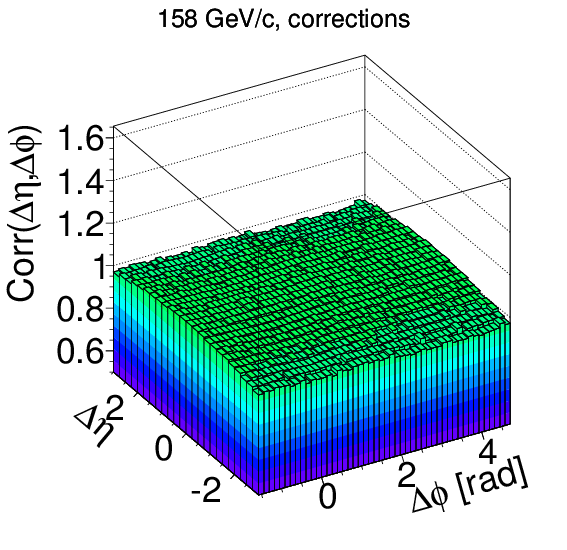}
  \includegraphics[width=0.25\textwidth]{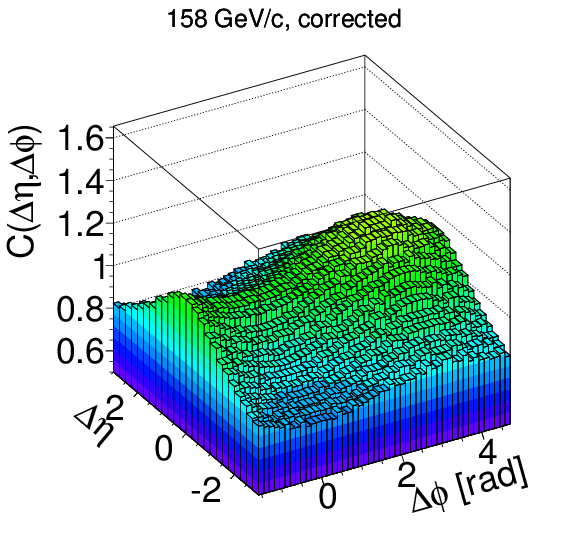}
  \caption{Examples of uncorrected correlation functions from data (left plots), correction factors (centre plots) and corrected two-particle correlation functions $C(\Delta\eta,\Delta\phi)$ (right plots) for all charge pairs. Results are for p+p interactions at 20~\GeVc (top) and 158~\GeVc (bottom).}
  \label{fig:corrections_example}
\end{figure*}

Examples of raw correlation functions, uncorrected for trigger bias, track selection cuts and reconstruction efficiency are presented in the left column of Fig.~\ref{fig:corrections_example} for all charge pairs.\footnote{Pairs of all, unlike-sign, positively, negatively charged particles are denoted by all, unlike-sign ($+-$), positive ($++$), negative ($--$) charge pairs for brief.}

\subsection{Corrections and uncertainties}

\subsubsection{Correction for reconstruction inefficiency and trigger bias}
\label{sec:det_eff_correction}

In order to correct the results for biases due to trigger and off-line event and track selection, detection efficiency, contribution of weak decays and secondary interaction products, the same analysis was also performed on simulated data. These effects, in general, may change the correlation function. In particular, weak decays of $\Lambda$ and $K^0_S$ hadrons lead to production of positively and negatively charged hadrons which are correlated by weak decay kinematics. The EPOS~\cite{Werner:2005jf} model was used for event generation since it provides a good description of the NA61/SHINE results on the yields of both non-strange and strange particles in p+p collisions~\cite{Abgrall:2013qoa,Aduszkiewicz:2015dmr,Pulawski:2015tka}. Correction factors $\text{Corr}(\Delta\eta,\Delta\phi)$ were calculated bin-by-bin as the ratio of the correlation functions for simulated events from the \Epos~\cite{Werner:2005jf} model (``pure'') and the same events after processing through \Geant \cite{Geant3} detector simulation and reconstruction (``rec''), and filtered using event and track selection cuts.

\begin{equation}
\label{eq:corrfactor}
  \text{Corr}(\Delta\eta,\Delta\phi) =
  \frac{C^{\text{sim}}_{\text{pure}}(\Delta\eta,\Delta\phi)}
  {C^{\text{sim}}_{\text{rec}}(\Delta\eta,\Delta\phi)},
\end{equation}
where $C^{\text{sim}}_{\text{pure}}$ and $C^{\text{sim}}_{\text{rec}}$ are the correlation function $C(\Delta\eta,\Delta\phi)$ obtained for simulated events before and after detector simulation and reconstruction. For both ``pure'' and ``rec'' events the \NASixtyOne acceptance filter was applied.

Corrected results $C(\Delta\eta,\Delta\phi)$ were obtained by multiplying the uncorrected correlation function by the corresponding correction, namely:

\begin{equation}
\label{eq:corrected}
  C(\Delta\eta,\Delta\phi) = C^{\text{raw}}(\Delta\eta,\Delta\phi)
     \cdot \text{Corr}(\Delta\eta,\Delta\phi).
\end{equation}

Correction factors $\text{Corr}(\Delta\eta,\Delta\phi)$ and corrected correlation functions $C(\Delta\eta,\Delta\phi)$ for all charge pairs in p+p interactions at 20 and 158~\GeVc are shown in Fig.~\ref{fig:corrections_example} as examples. The values of the correction factors range between 0.9 and 1.1.

\subsubsection{Statistical uncertainties}
\label{sec:statistical_errors}

Statistical uncertainties of the correlation function are calculated as:
\begin{equation}
  \label{eq:sigma_C}
  \sigma^2(C) = \sqrt{\left [\text{Corr} \cdot \sigma(C^{\text{raw}}) \right]^2
    + \left[ C^{\text{raw}} \cdot \sigma(\text{Corr}) \right]^2}
\end{equation}
in each $(\Delta\eta,\Delta\phi)$ bin. Statistical uncertainties are found to be approximately independent of $\Delta\phi$ but increase significantly with increasing $\Delta\eta$. Depending on the beam momentum and charge combination the statistical uncertainties of the $C$ function in individual ($\Delta\eta,\Delta\phi$) bins are about 5\% for $\Delta\eta \approx 0$ (all charge pairs at top beam momentum) to more than 20\% for $\Delta\eta \approx 3$ (negatively charge pairs at the lowest beam momentum).

\subsubsection{Estimation of systematic uncertainties}
\label{sec:systematic_errors}

In order to estimate systematic uncertainties the data were analysed with loose and tight event and track selection cuts. By changing the cuts, one changes the magnitude of the corrections due to various biasing effects. In case of simulation perfectly reproducing the data, corrected results should be independent of the cuts. A dependence on cuts is due to imperfections of the simulation and is used as an estimate of the systematic uncertainty. For example, systematic uncertainty caused by weakly decaying particles is estimated by varying $B_x$ and $B_y$ cuts. The standard set of cuts was presented in Sec.~\ref{sec:selection} and is tabulated in Table~\ref{tab:standard_loose_tight} together with loose and tight cuts. 

\begin{table}
  \centering
  \begin{tabular}{|l|c|c|c|}
    \hline
    \multicolumn{4}{|c|}{\textbf{Event cuts}} \\
    \hline
    & \textbf{Loose} & \textbf{Standard} & \textbf{Tight} \\
    \hline
    \textbf{Event with interaction trigger} & \multicolumn{3}{c|}{applied} \\
    \hline
    \textbf{BPD} & \multicolumn{3}{c|}{applied} \\
    \hline
    \textbf{No off-time beam particles} & disabled & $< \pm 1\mu$s & $< \pm 5\mu$s \\
    \hline
    \textbf{At least one track in TPCs} & \multicolumn{3}{c|}{applied} \\
    \hline
    \textbf{Vertex position $z$} & $\pm 11$~cm & $\pm 10$~cm & $\pm 7$~cm \\
    \hline
    \textbf{Elastic event} & \multicolumn{3}{c|}{applied} \\
    \hline
    \multicolumn{4}{|c|}{\textbf{Track cuts}} \\
    \hline
    \textbf{Charge $\neq 0$} & \multicolumn{3}{c|}{applied} \\
    \hline
    \textbf{Total TPC points} & $\geq 10$ & \multicolumn{2}{c|}{$\geq 30$} \\
    \hline
    \textbf{VTPC (GTPC) points} & $> 10(5)$ & $\geq 15(5)$ & $\geq 30(6)$ \\
    \hline
    \textbf{$|B_x|$} & $\leq 5$~cm & $\leq 4$~cm & $\leq 1$~cm \\
    \hline
    \textbf{$|B_y|$} & $\leq 2.5$~cm & $\leq 2$~cm & $\leq 0.5$~cm \\
    \hline
    \textbf{$p_T$ cut} & \multicolumn{3}{c|}{applied} \\
    \hline
    \textbf{$e^- e^+$ cut} & \multicolumn{3}{c|}{applied} \\
    \hline
  \end{tabular}
  \caption{Event (top) and track (bottom) selection cuts. The standard cuts (centre) are used to obtain the final results, whereas the loose (left) and tight (right) cuts are employed to estimate systematic uncertainties (see Sec.~\ref{sec:event_cuts} and~\ref{sec:track_cuts}, respectively).}
  \label{tab:standard_loose_tight}
\end{table}

Results for both sets of cuts were subtracted bin-by-bin ($\text{loose} - \text{tight}$). Since the differences in all bins follow Gaussian distributions with mean close to 0, the systematic uncertainties were estimated as the standard deviation of the distribution. The procedure was performed for all charge combinations (all charge, unlike-sign, positively and negatively charge pairs) and at 20, 40 and 158~\GeVc. The systematic uncertainties are generally below 1\%. The exception are correlation functions at 20~\GeVc at higher $\Delta\eta$, where the systematic uncertainties are about 5\%.


\section{Results and discussion}\label{sec:results}

\subsection{Two-particle correlation function $C(\Delta\eta,\Delta\phi)$}

\begin{figure*}
  \centering
  \includegraphics[width=0.25\textwidth]{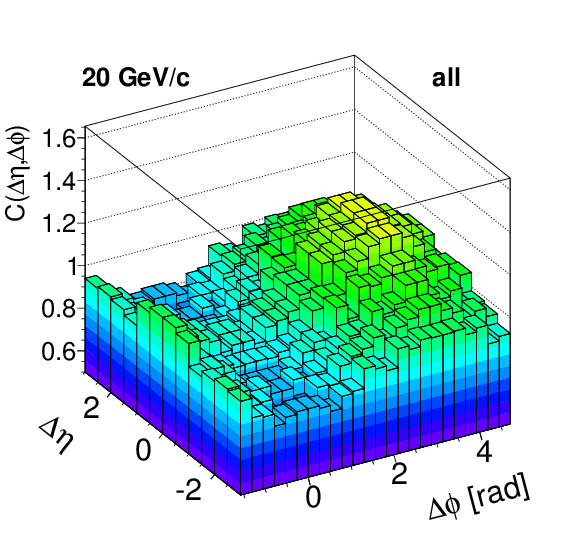}
  \includegraphics[width=0.25\textwidth]{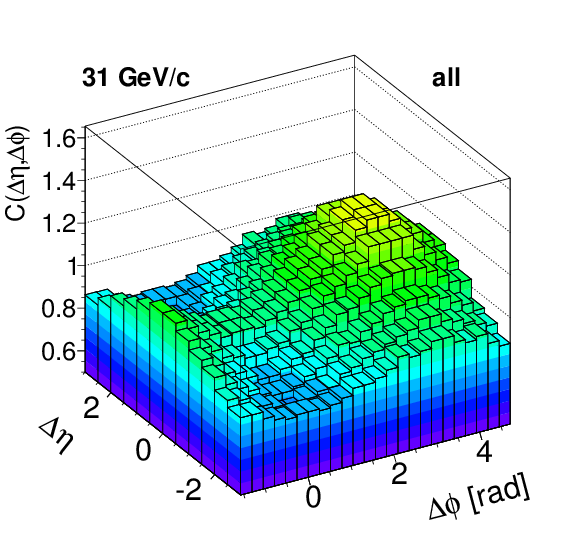}
  \includegraphics[width=0.25\textwidth]{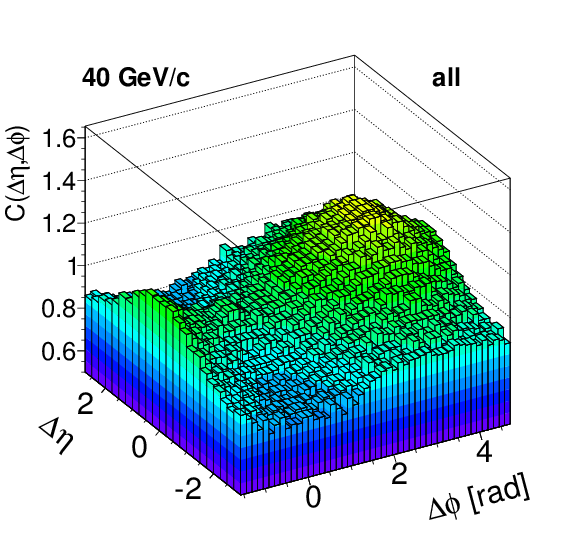}
  \\
  \includegraphics[width=0.25\textwidth]{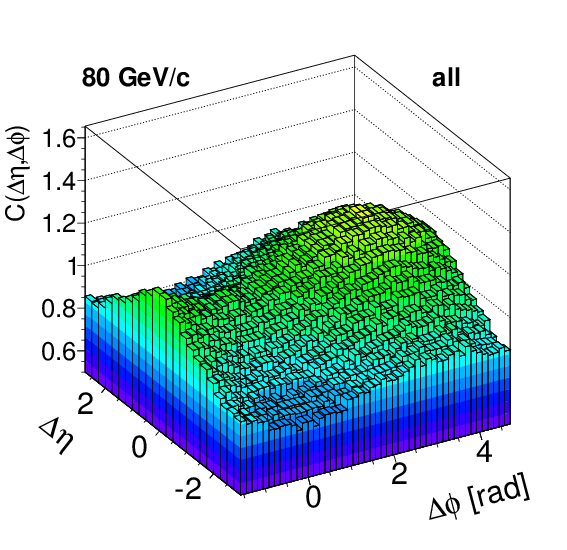}
  \includegraphics[width=0.25\textwidth]{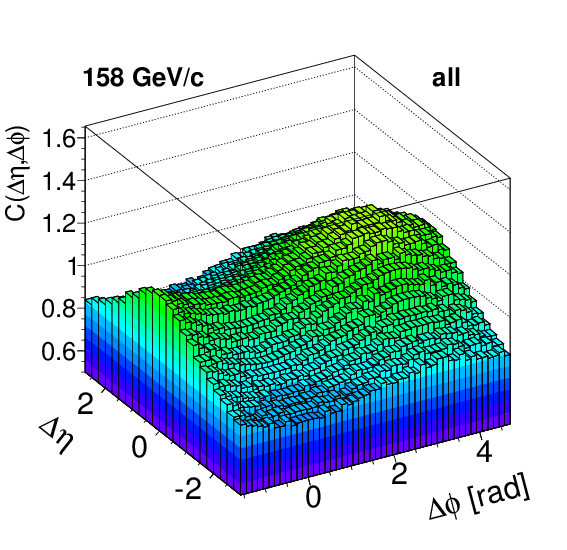}
  \caption{Two-particle correlation function $C(\Delta\eta,\Delta\phi)$ for all charge pairs in inelastic p+p interactions at 20-158~\GeVc. The correlation function is mirrored around $(\Delta\eta,\Delta\phi)=(0,0)$.}
  \label{fig:data_corr_all}
\end{figure*}

\begin{figure*}
  \centering
  \includegraphics[width=0.25\textwidth]{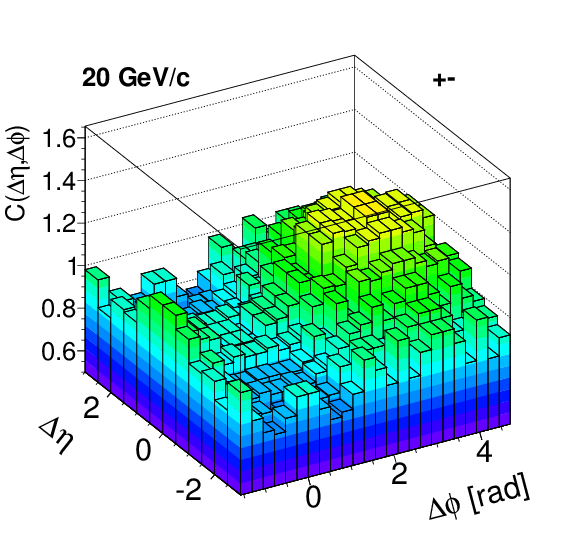}
  \includegraphics[width=0.25\textwidth]{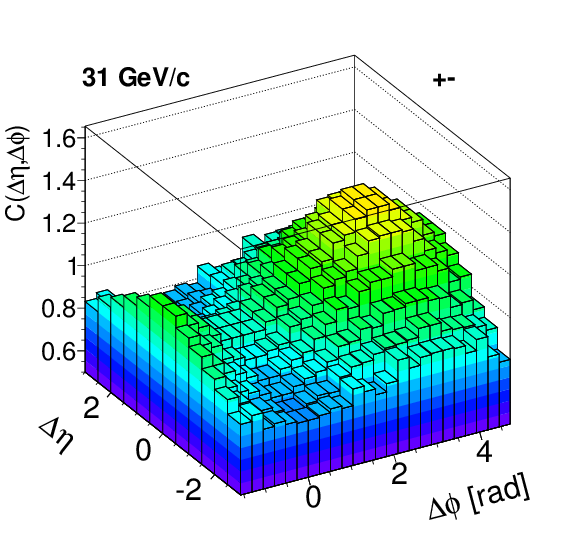}
  \includegraphics[width=0.25\textwidth]{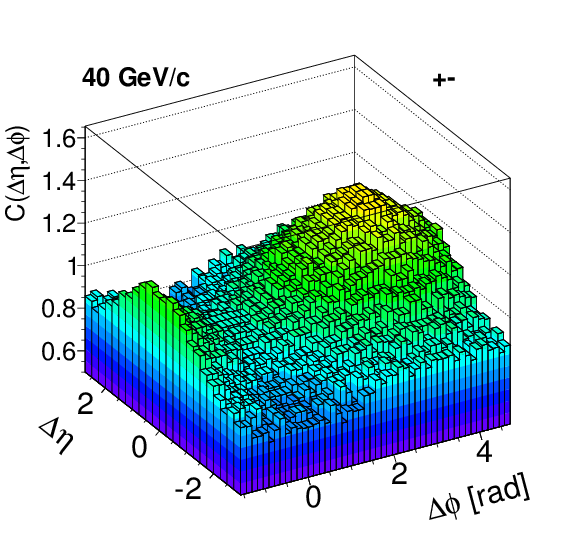}
  \\
  \includegraphics[width=0.25\textwidth]{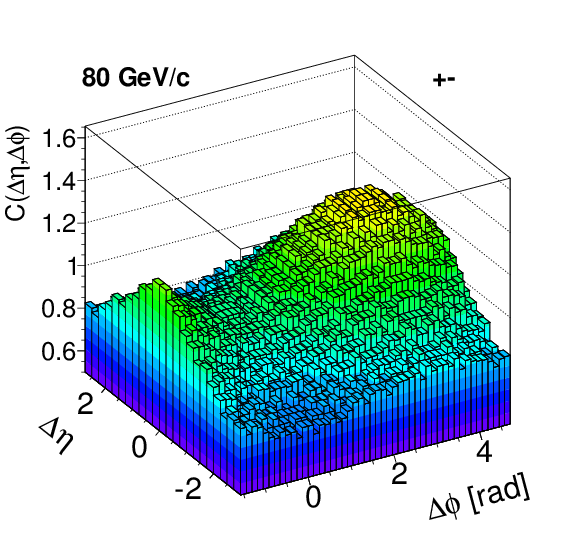}
  \includegraphics[width=0.25\textwidth]{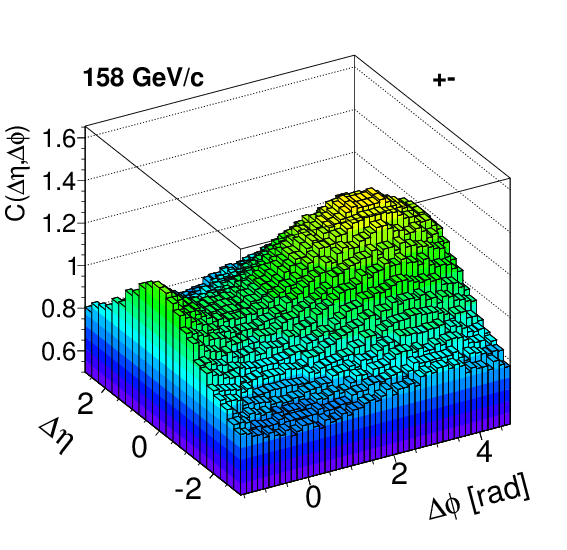}
  \caption{Two-particle correlation function $C(\Delta\eta,\Delta\phi)$ for unlike-sign pairs in inelastic p+p interactions at 20-158~\GeVc. The correlation function is mirrored around $(\Delta\eta,\Delta\phi)=(0,0)$.}
  \label{fig:data_corr_unlike}
\end{figure*}

\begin{figure*}
  \centering
  \includegraphics[width=0.25\textwidth]{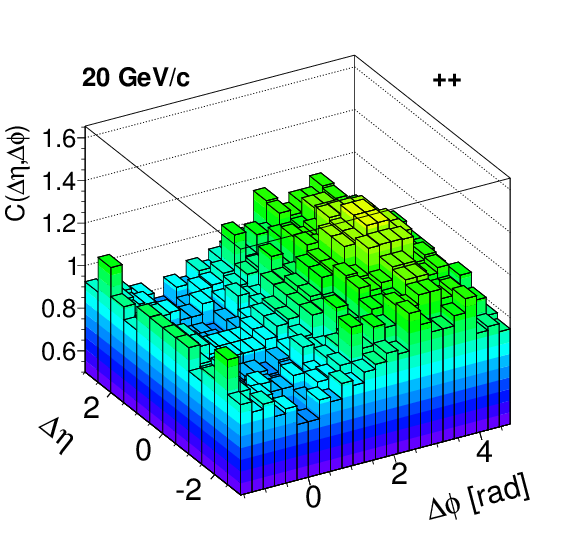}
  \includegraphics[width=0.25\textwidth]{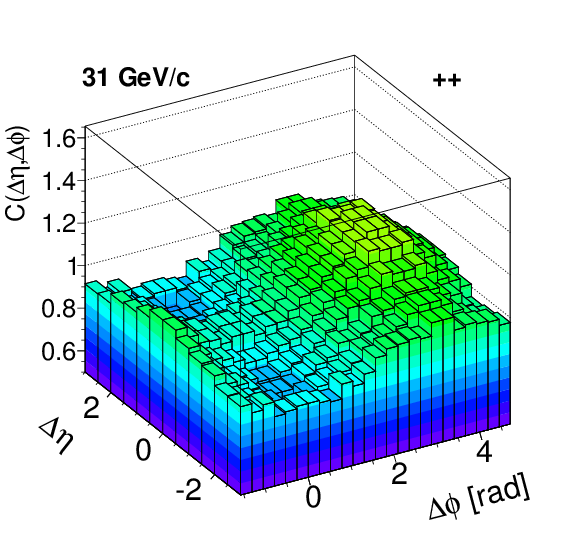}
  \includegraphics[width=0.25\textwidth]{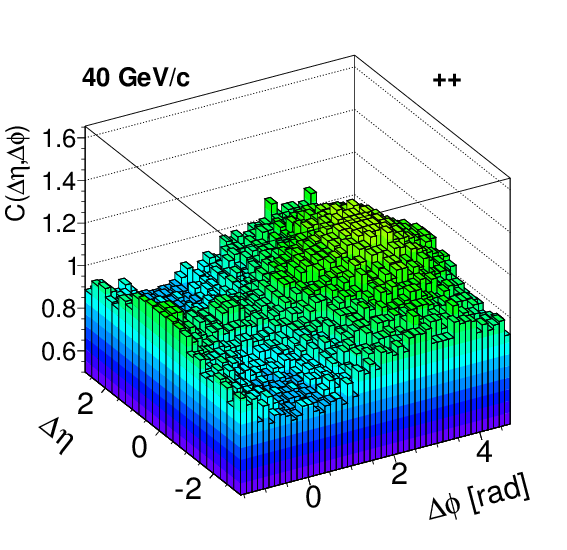}
  \\
  \includegraphics[width=0.25\textwidth]{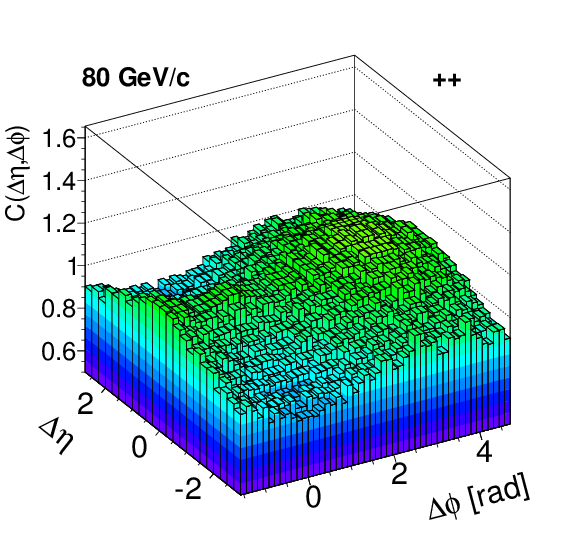}
  \includegraphics[width=0.25\textwidth]{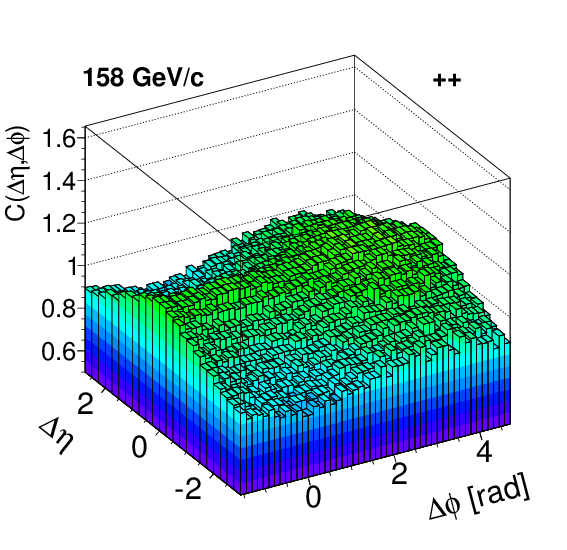}
  \caption{Two-particle correlation function $C(\Delta\eta,\Delta\phi)$ for positive charge pairs in inelastic p+p interactions at 20-158~\GeVc. The correlation function is mirrored around $(\Delta\eta,\Delta\phi)=(0,0)$.}
  \label{fig:data_corr_pos}
\end{figure*}

\begin{figure*}
  \centering
  \includegraphics[width=0.25\textwidth]{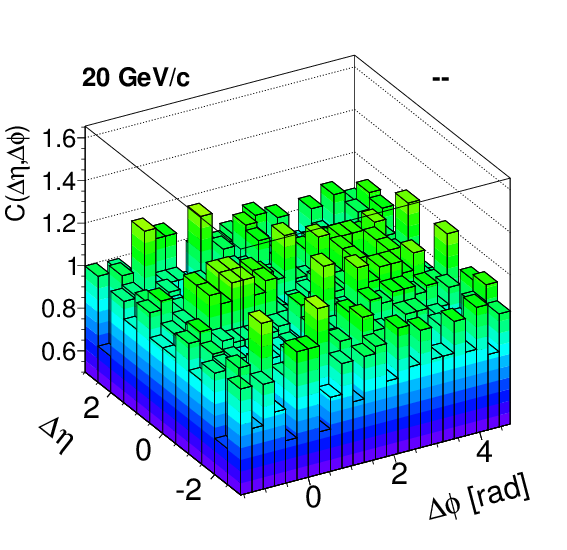}
  \includegraphics[width=0.25\textwidth]{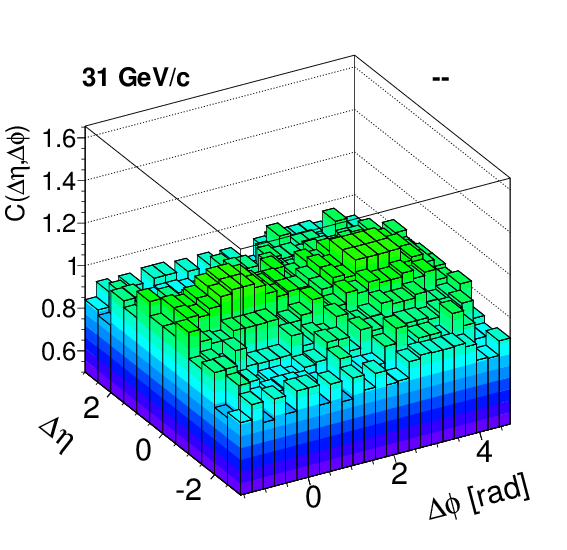}
  \includegraphics[width=0.25\textwidth]{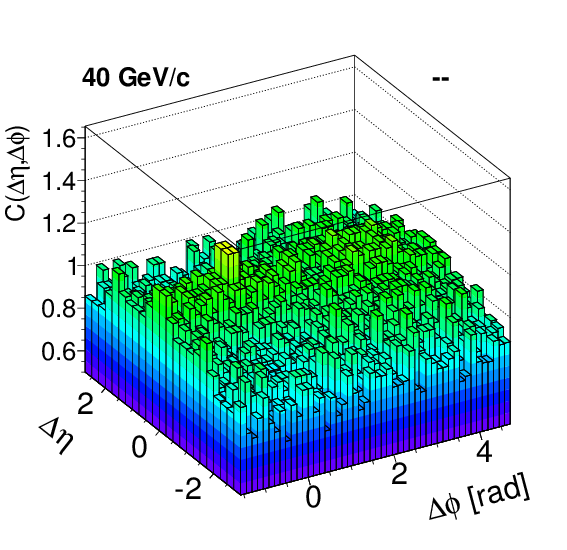}
  \\
  \includegraphics[width=0.25\textwidth]{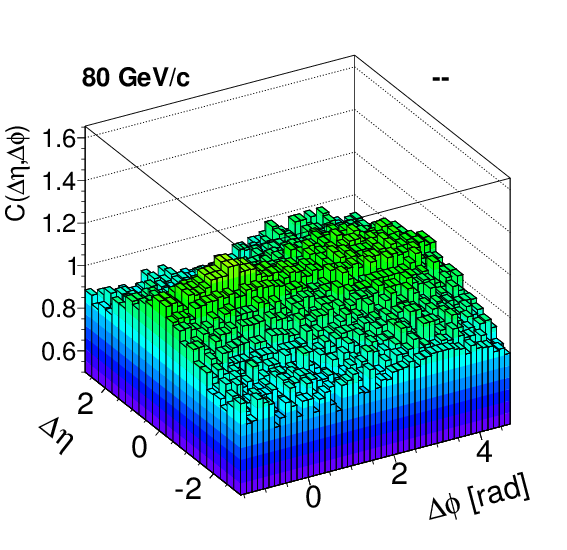}
  \includegraphics[width=0.25\textwidth]{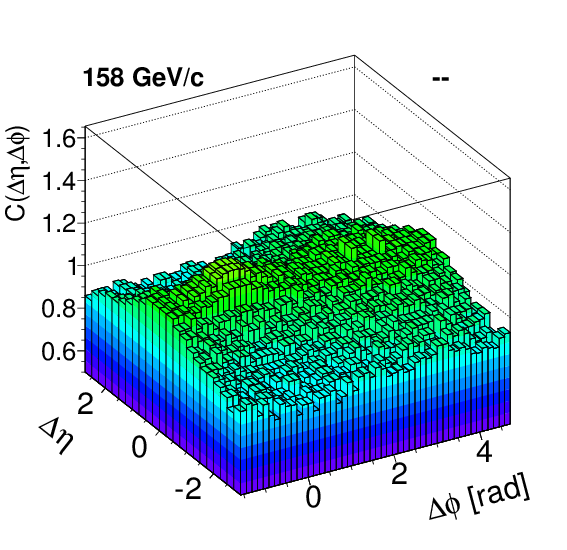}
  \caption{Two-particle correlation function $C(\Delta\eta,\Delta\phi)$ for negative charge pairs in inelastic p+p interactions at 20-158~\GeVc. The correlation function is mirrored around $(\Delta\eta,\Delta\phi)=(0,0)$.}
  \label{fig:data_corr_neg}
\end{figure*}

The corrected correlation functions for all charge pair combinations (all charge pairs, unlike-sign pairs, positively and negatively charge pairs) are presented in Figs.~\ref{fig:data_corr_all}, \ref{fig:data_corr_unlike}, \ref{fig:data_corr_pos} and \ref{fig:data_corr_neg}, respectively. Their values lie in the range between 0.8 and 1.4. Vanishing two-particle correlations result in $C = 1$. 

Statistical and systematic uncertainties were calculated using the procedures described in Secs.~\ref{sec:statistical_errors} and~\ref{sec:systematic_errors}. The numerical values of the correlation functions and of statistical and systematic uncertainties are available in Ref.~\cite{DetaDphiEDMSresults}.

The main features of the results are:
\begin{enumerate}[(i)]
\item A maximum at $(\Delta\eta,\Delta\phi) \approx (0,\pi)$, most prominent for all charge and unlike-sign pairs and significantly weaker for like-sign pairs. The most probable explanation is the contribution from resonance decays, mostly from abundantly produced $\rho^0 \rightarrow \pi^+ + \pi^-$. The weaker maximum for positive charge pairs can be attributed to e.g. $\Delta^{++}$ resonance decay. No such maximum is observed for negative charge pairs consistent with the fact that there are almost no resonances decaying into two or more negative charge pairs.
\item A $\cos(\Delta\phi)$ modulation appearing as a minimum near $\Delta\phi = 0$ and maximum near $\Delta\phi = \pi$ at all values of $\Delta\eta$ for all combinations of charges. Stronger in all charge and unlike-sign pairs, weaker but still visible in positive charge and barely noticeable for negative charge pairs. The structure is probably due to momentum conservation.
\item A Gaussian-like enhancement around $\Delta\eta = 0$ along the full $\Delta\phi$ range, clearly visible for all charge and unlike-sign pairs, however, significantly weaker, but still noticeable in like-sign pairs. This feature may be connected with string fragmentation or flux-tube fragmentation~\cite{Wong:2015nwa}.
\item A hill around $(\Delta\eta,\Delta\phi) = (0,0)$ for like-sign pairs. For positive charged pairs it grows with increasing beam momentum, but it is independent of beam momentum for negative charge pairs. Since the products of $\gamma$ decay were rejected during the analysis, the hill is probably caused by Bose--Einstein statistics.
\item Clearly, there is no jet-like peak at $(\Delta\eta,\Delta\phi) = (0,0)$. This indicates that the contribution of hard scattering effects is small as expected from their small probability at SPS energies (omitting the cut $p_T < 1.5$~\GeVc was tried and did not change the results).
\end{enumerate}

\subsection{Projections onto $C(\Delta\eta)$ and $C(\Delta\phi)$}

\begin{figure*}
  \centering
  \includegraphics[width=\textwidth]{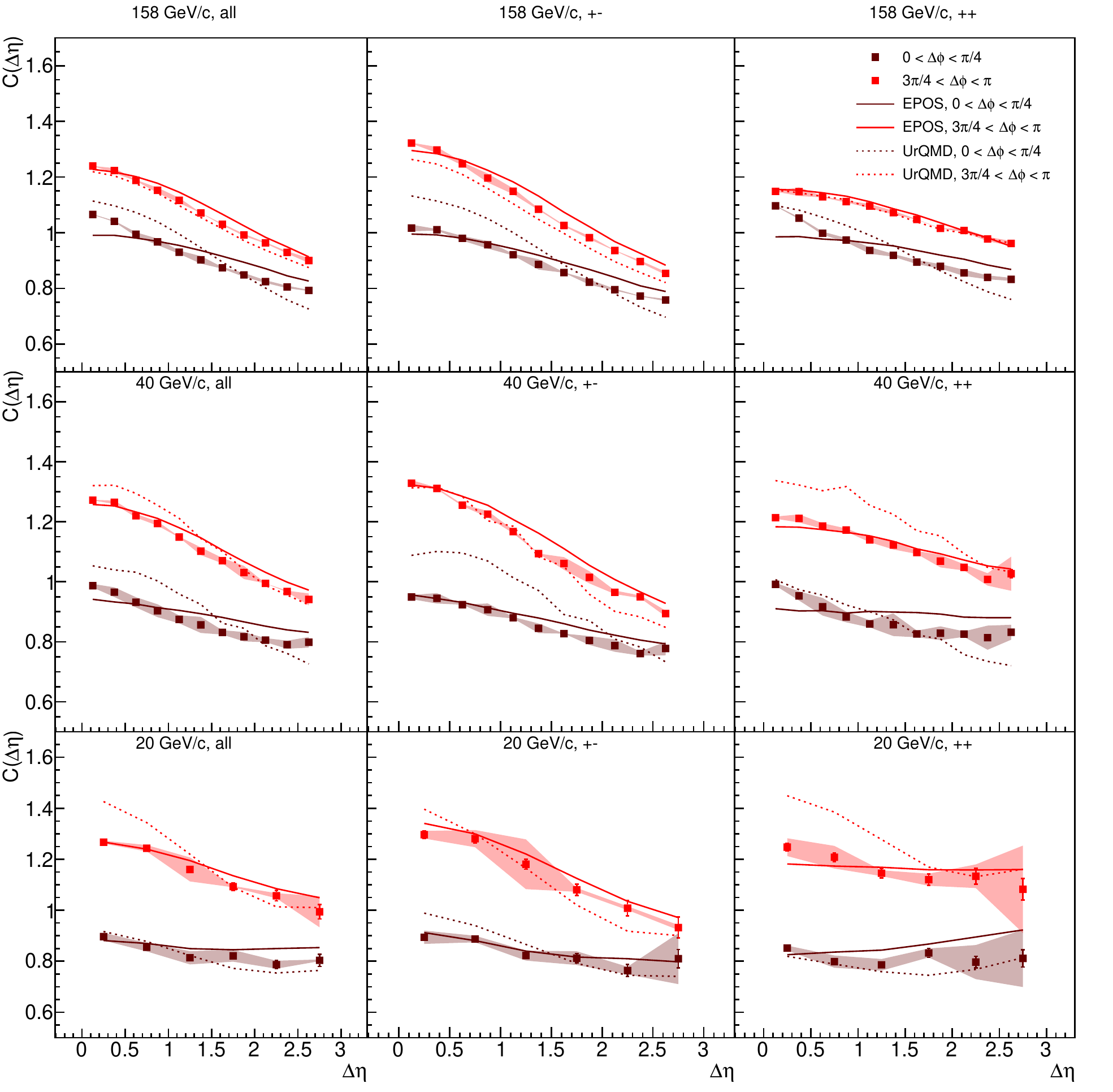}
  \caption{Two-particle correlation function $C(\Delta\eta)$ obtained from projection of $C(\Delta\eta,\Delta\phi)$ onto the $\Delta\eta$ axis for two subranges of $\Delta\phi$. Left column shows the results for all charge pairs, middle column -- unlike-sign pairs, right column -- positive charge pairs. Vertical bars denote statistical and shaded regions denote systematic uncertainties. Predictions of the \Epos model are shown by solid curves and the UrQMD model by dotted curves. Legend applies to all panels.}
  \label{fig:deta_projections}
\end{figure*}

To make the results easier to compare quantitatively, slices of the two-dimensional correlation function are also presented. The $C(\Delta\eta)$ function was obtained by projecting $C(\Delta\eta,\Delta\phi)$ onto $\Delta\eta$ in the $\Delta\phi$ intervals: $0 < \Delta\phi < \pi/4$ and $3\pi/4 < \Delta\phi < \pi$. The results for 20, 40, and 158~\GeVc beam momenta and all charge, unlike-sign and positive charged pairs, together with uncertainties as well as predictions of the \Epos and UrQMD models are presented in Fig.~\ref{fig:deta_projections}. Note that due to the projection the statistical uncertainties of the projected correlation function are considerably smaller than for individual two-dimensional bins.

\begin{figure*}
  \centering
  \includegraphics[width=\textwidth]{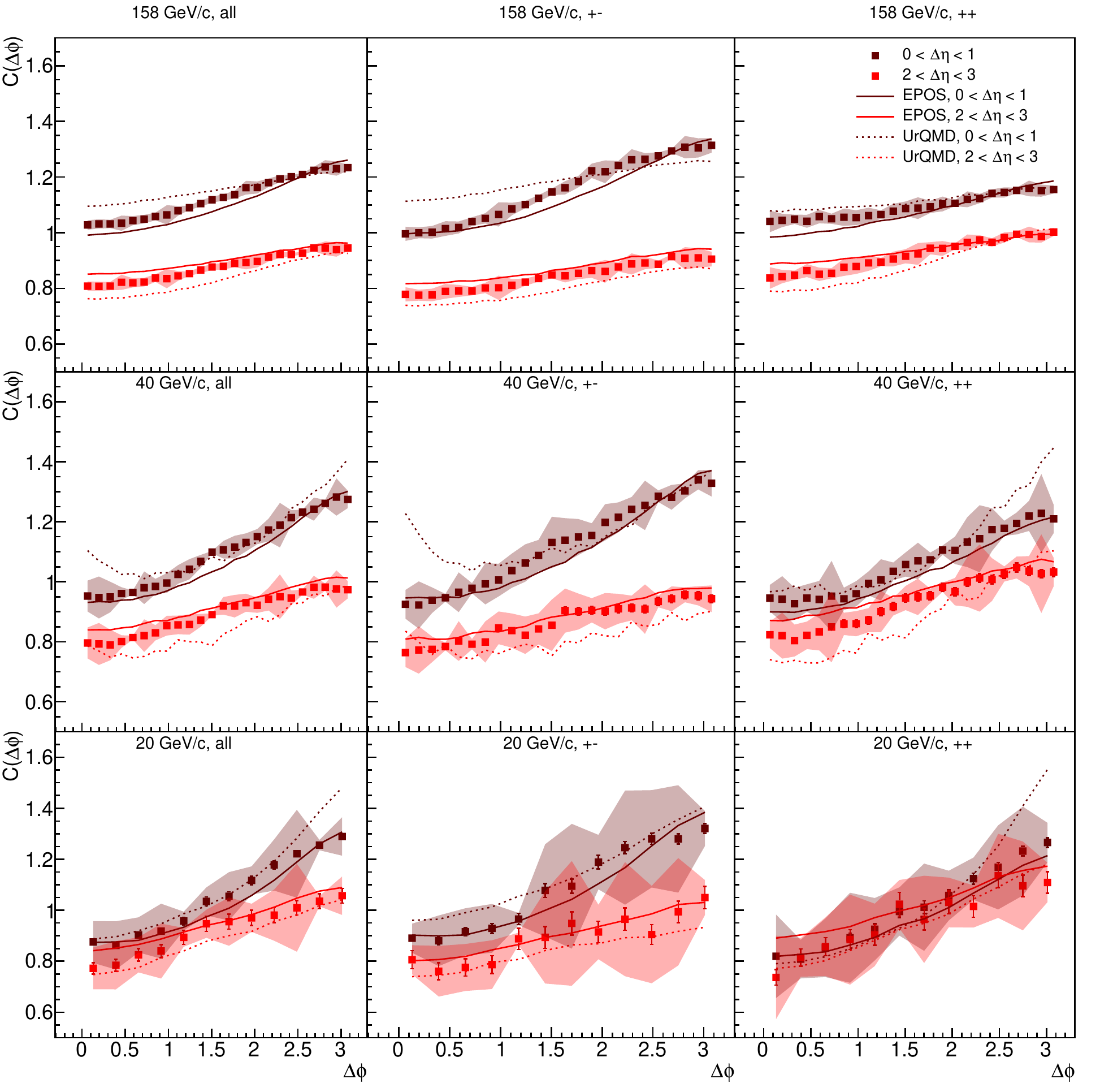}
  \caption{Two-particle correlation function $C(\Delta\phi)$ obtained from projection of $C(\Delta\eta,\Delta\phi)$ onto the $\Delta\phi$ axis for two subranges of $\Delta\eta$. Left column shows the results for all charge pairs, middle column -- unlike-sign pairs, right column -- positive charge pairs. Vertical bars denote statistical and shaded regions denote systematic uncertainties. Predictions of the \Epos model are shown by solid curves and the UrQMD model by dotted curves. Legend applies to all panels.}
  \label{fig:dphi_projections}
\end{figure*}

The two-particle correlation function $C(\Delta\phi)$ was obtained from projection of $C(\Delta\eta,\Delta\phi)$ onto the $\Delta\phi$ axis in two $\Delta\eta$ intervals: $0 < \Delta\eta < 1$ and $2 < \Delta\eta < 3$. The results are presented in Fig.~\ref{fig:dphi_projections}. The tendencies shown in Figs.~\ref{fig:deta_projections} and~\ref{fig:dphi_projections} are reproduced well by the \Epos model.

\subsection{Acceptance effects}
\label{sec:acceptance_effects}

\begin{figure*}
  \centering
    \includegraphics[width=0.25\textwidth]{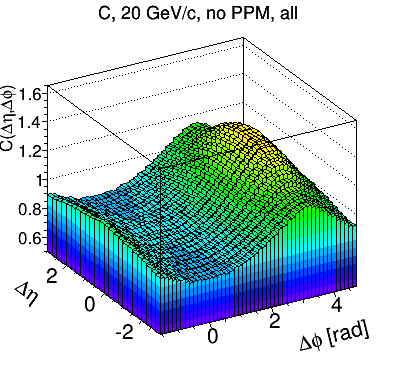}
    \includegraphics[width=0.25\textwidth]{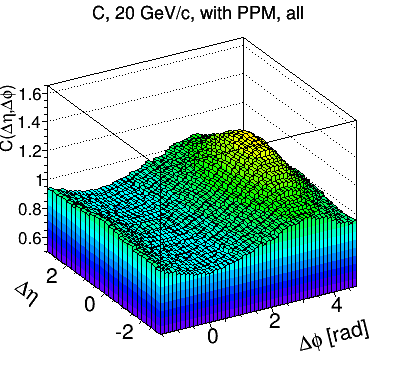}
    \includegraphics[width=0.25\textwidth]{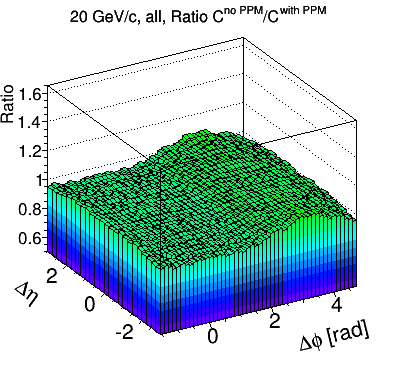}
    \\
    \includegraphics[width=0.25\textwidth]{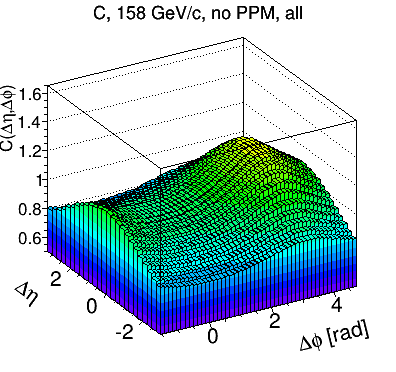}
    \includegraphics[width=0.25\textwidth]{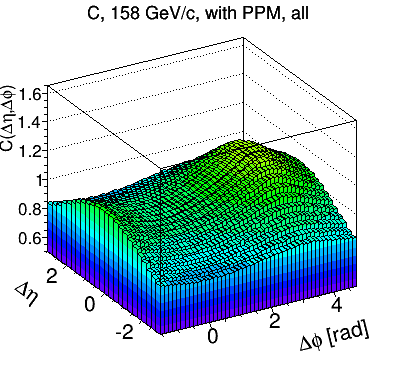}
    \includegraphics[width=0.25\textwidth]{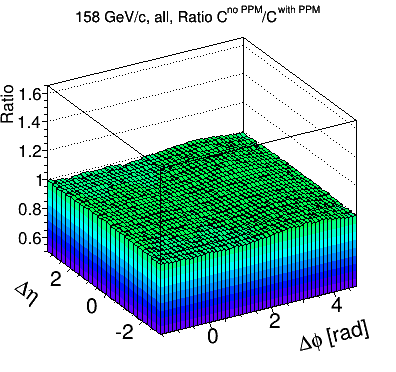}
  \caption{\NASixtyOne detector acceptance effects studied with the \Epos~\cite{Werner:2005jf} model. Left plots show correlation functions in almost complete acceptance (only $p_T < 1.5$~\GeVc cut was applied), middle plots show correlation functions in \NASixtyOne acceptance, right plots -- ratios of these two. Top row shows results for 20~\GeVc, bottom row for 158~\GeVc.}
  \label{fig:acc_effects}
\end{figure*}

Acceptance effects were studied using events generated with the \Epos model. Namely, correlation functions $C(\Delta\eta,\Delta\phi)$ were produced for \Epos events in almost complete acceptance (full range of pseudorapidity and azimuthal angle but $p_T < 1.5$~\GeVc) and the \NASixtyOne detector acceptance (with application of the Particle Population Matrix~\cite{ppm_edms}). The results (Fig.~\ref{fig:acc_effects}) show that the correlation functions for both acceptances are qualitatively similar (only the enhancement close to $\Delta\phi \approx \pi$ are 3-7\% stronger in case of plots with larger acceptance). Thus, although the \NASixtyOne acceptance is limited, for the \Epos model, similar physics conclusions can be drawn as from almost complete acceptance plots. The right panels of Fig.~\ref{fig:acc_effects} display the ratios of two correlation functions calculated with both acceptances. The effects of the limited \NASixtyOne acceptance are stronger at 20~\GeVc. It is essential to apply the \NASixtyOne acceptance filter to model calculations before comparing to the measurements.

\section{Comparison with models}\label{sec:models}

\subsection{Comparison with the \Epos and UrQMD models}

\begin{figure*}
  \centering
  \includegraphics[width=0.24\textwidth]{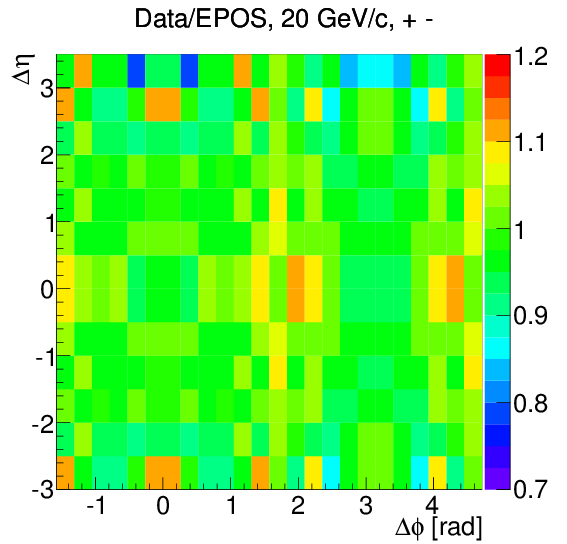}
  \includegraphics[width=0.24\textwidth]{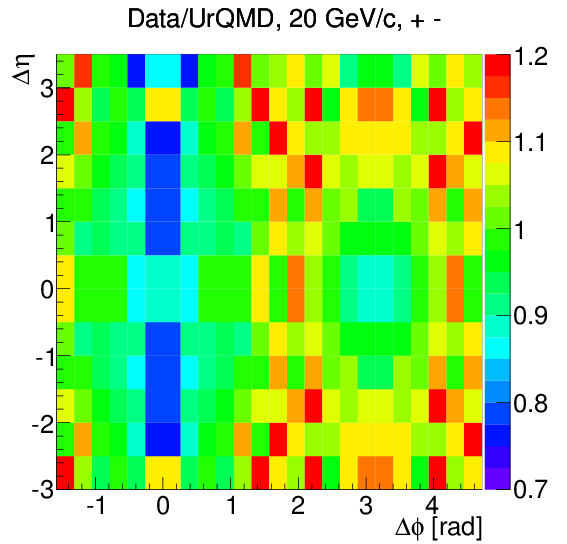}
  \includegraphics[width=0.24\textwidth]{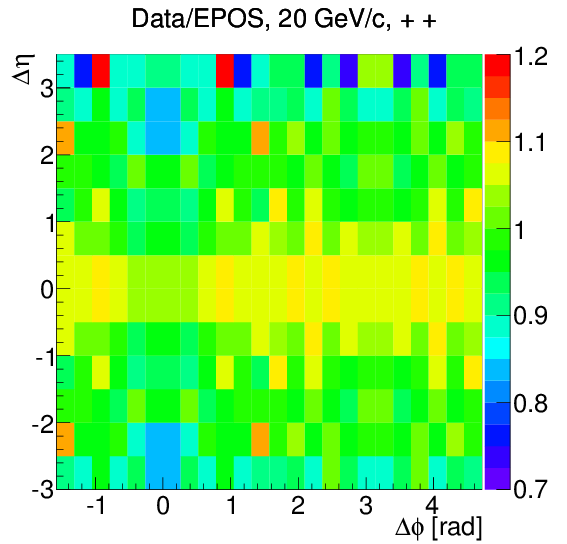}
  \includegraphics[width=0.24\textwidth]{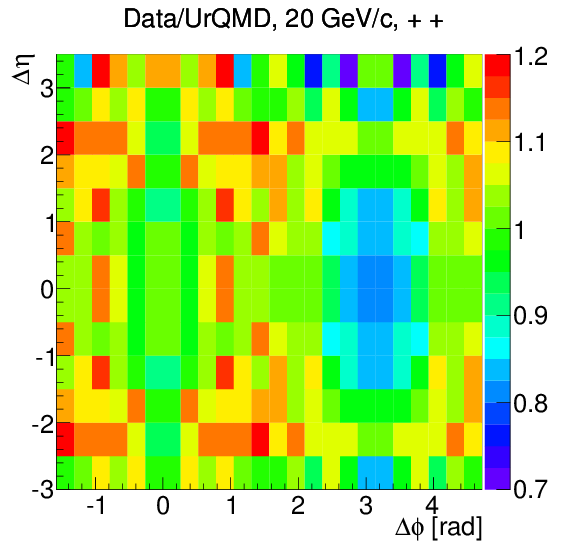}
  \\
  \includegraphics[width=0.24\textwidth]{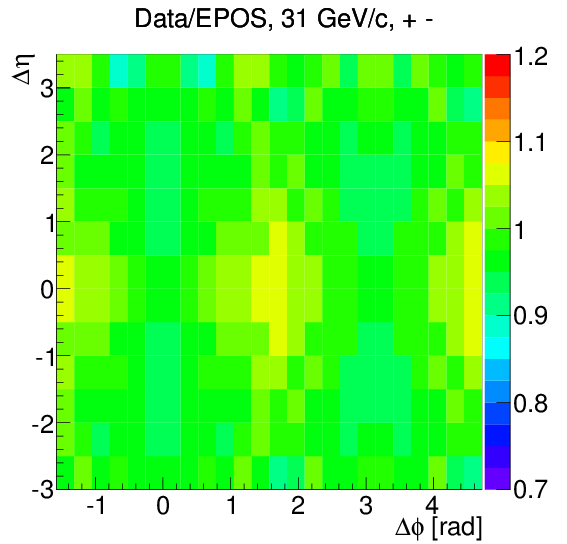}
  \includegraphics[width=0.24\textwidth]{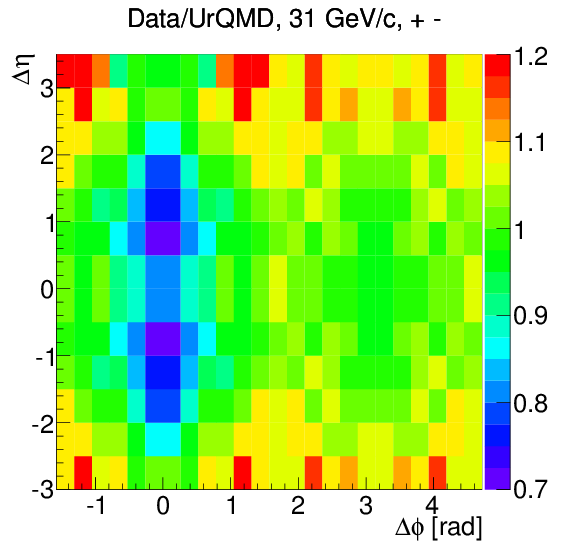}
  \includegraphics[width=0.24\textwidth]{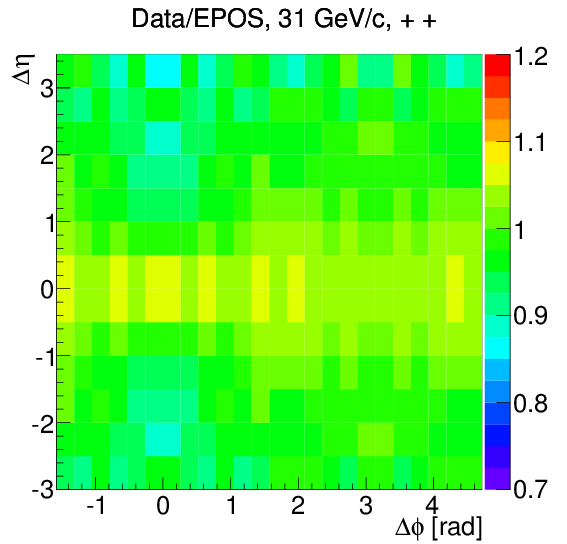}
  \includegraphics[width=0.24\textwidth]{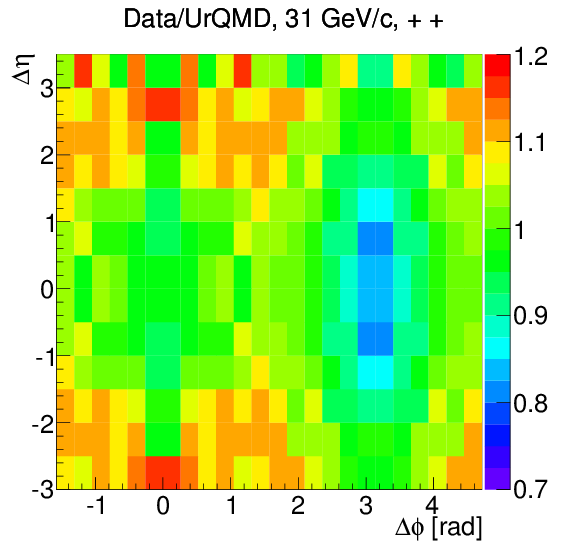}
  \\
  \includegraphics[width=0.24\textwidth]{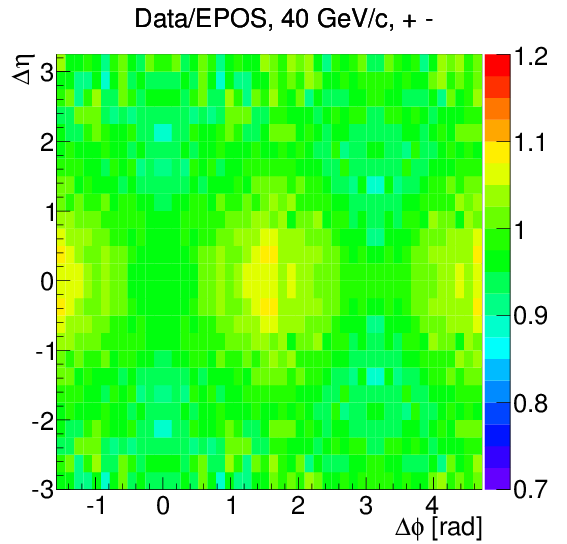}
  \includegraphics[width=0.24\textwidth]{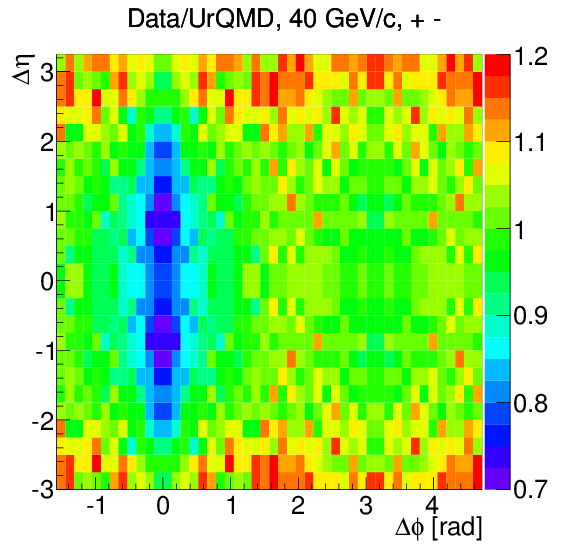}
  \includegraphics[width=0.24\textwidth]{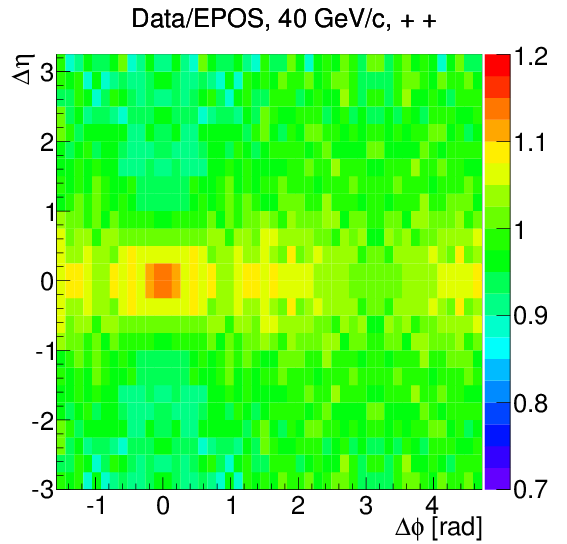}
  \includegraphics[width=0.24\textwidth]{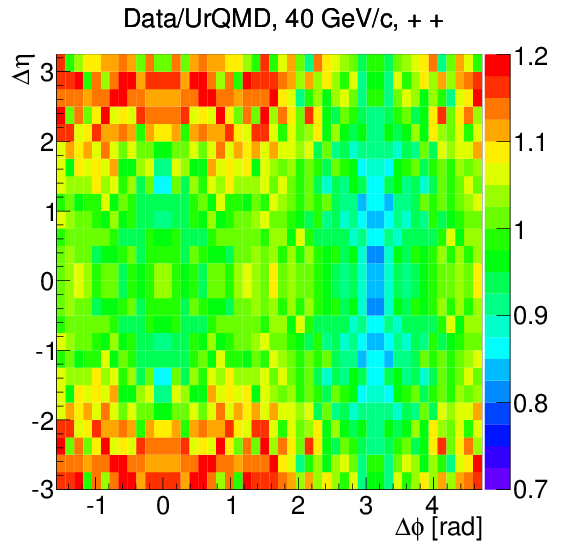}
  \\
  \includegraphics[width=0.24\textwidth]{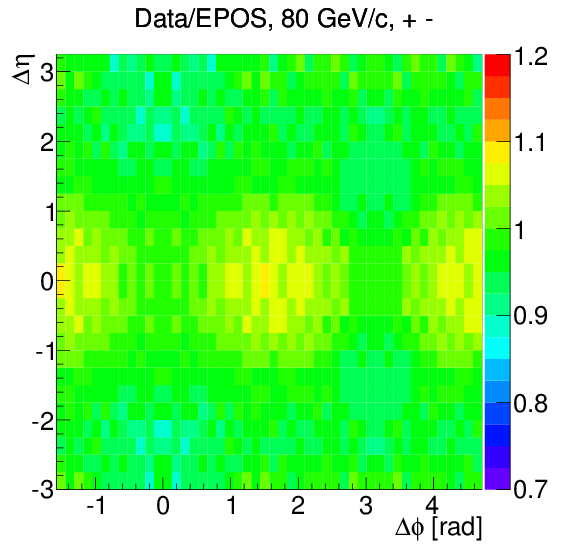}
  \includegraphics[width=0.24\textwidth]{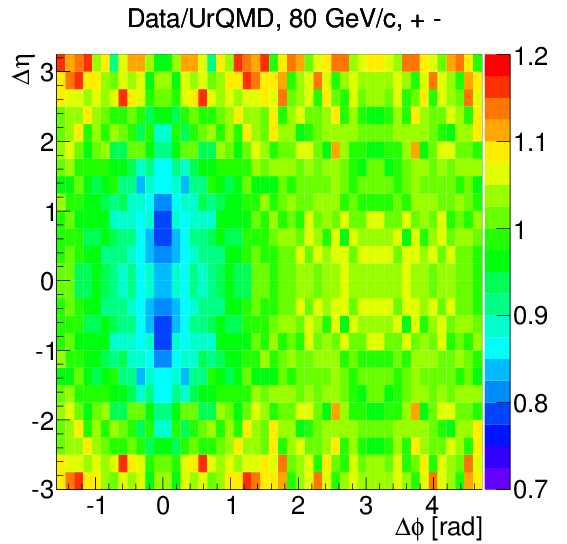}
  \includegraphics[width=0.24\textwidth]{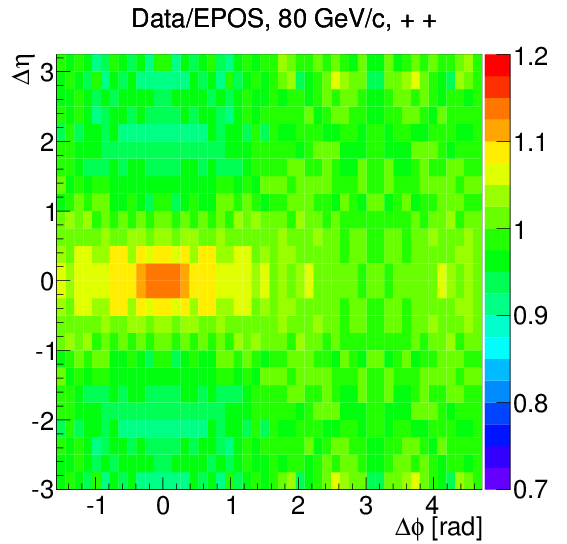}
  \includegraphics[width=0.24\textwidth]{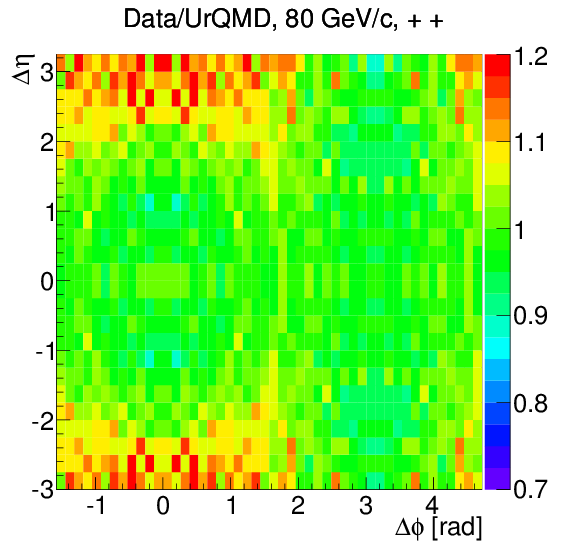}
  \\
  \includegraphics[width=0.24\textwidth]{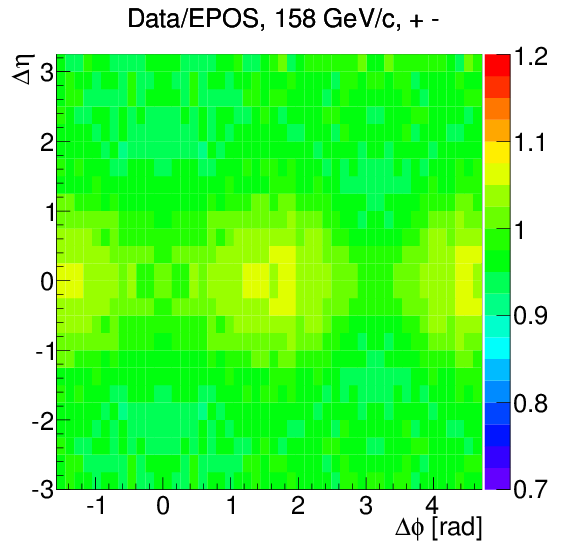}
  \includegraphics[width=0.24\textwidth]{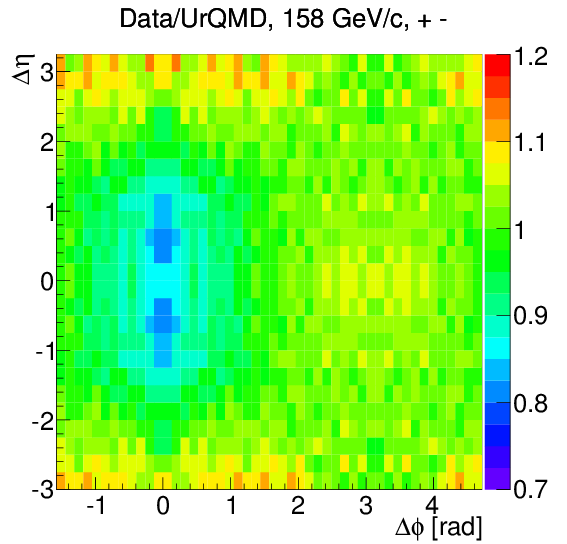}
  \includegraphics[width=0.24\textwidth]{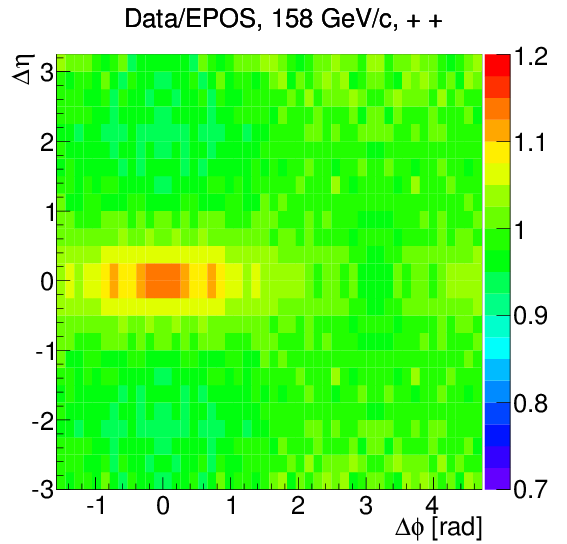}
  \includegraphics[width=0.24\textwidth]{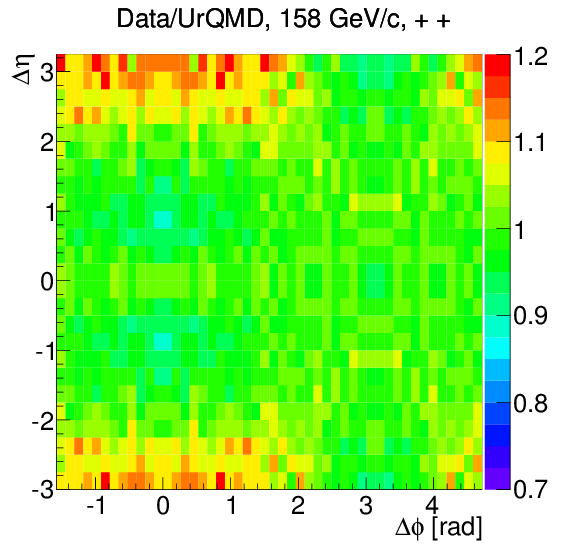}
  \caption{The ratio of correlation functions $C(\Delta\eta,\Delta\phi)$ for data and models. The two left columns show results for unlike-sign pairs and the two right columns results for positive charge pairs. Each row shows results for one beam momentum.}
  \label{fig:data_vs_mc_ratio}
\end{figure*}

In order to compare the experimental results and predictions of the \Epos~1.99~\cite{Werner:2005jf} and UrQMD~3.4~\cite{Bass:1998ca,Bleicher:1999xi} models the ratio of correlation functions obtained from data and models was calculated. The same acceptance was used for the \NASixtyOne data and for the \Epos and UrQMD models (see Ref.~\cite{ppm_edms}). The results for unlike-sign and positive charge pairs are shown in Fig.~\ref{fig:data_vs_mc_ratio}.
In general, \Epos~1.99 reproduces the experimental results from p+p interactions well (see first and third column in Fig.~\ref{fig:data_vs_mc_ratio}). However, the model does not reproduce the peak of about 20\% at $(\Delta\eta,\Delta\phi)=(0,0)$ seen for positive charge pairs (third column in Fig.~\ref{fig:data_vs_mc_ratio}) at 40, 80 and 158~\GeVc. This demonstrates an excess of the number of real data over model pairs in that region. Note that the \Epos model does not include either Bose--Einstein correlations or Coulomb interactions which are expected to produce this kind of correlation. Moreover, a weak enhancement of the ratio of about 10\% is seen near $\Delta\eta \approx 0$ and $\Delta\phi \approx \pi \text{ and } \frac{3}{4}\pi$ (first column in Fig.~\ref{fig:data_vs_mc_ratio}). This suggests a slight underestimate of the correlation in the valleys of the $\cos(\Delta\phi)$ modulation in \Epos.

The ratio of the experimental results and the predictions of the UrQMD~3.4 model is shown in the second and fourth column of Fig.~\ref{fig:data_vs_mc_ratio}. The UrQMD model reproduces data less well than \Epos. The most visible discrepancy is a $\Delta\eta$-wide suppression of the ratio in unlike-sign pairs (second column in Fig.~\ref{fig:data_vs_mc_ratio}) which suggests that UrQMD generates 25--30\% stronger long-range near-side ($\Delta\phi \approx 0$) correlations than existing in the data. Another inconsistency between data and model results can be seen in positive charge correlations (fourth column in Fig.\ref{fig:data_vs_mc_ratio}) where a suppression of up to 25\% is present on the away-side ($\Delta\phi \approx \pi$), strongest at the three lowest beam momenta, as well as an enhancement of more than 10\% at large $\Delta\eta$ on the same side in azimuthal angle.


\section{Summary}\label{sec:summary}

Two-particle correlations of charged particles in azimuthal angle and pseudorapidity were measured in inelastic p+p collisions at 20--158~\GeVc by the NA61/SHINE experiment at the CERN SPS. The results show structures which can be connected to phenomena such as resonance decays, momentum conservation and Bose--Einstein correlations. A comparison with the \Epos and UrQMD models was performed. The predictions of the \Epos model are close to the \NASixtyOne data whereas the results from UrQMD show larger discrepancies.


\section*{Acknowledgements}
We would like to thank the CERN EP, BE and EN Departments for the
strong support of NA61/SHINE.

This work was supported by the Hungarian Scientific Research Fund
(Grants OTKA 68506 and 71989), the J\'anos Bolyai Research Scholarship
of the Hungarian Academy of Sciences, the Polish Ministry of Science
and Higher Education (Grants 667\slash N-CERN\slash2010\slash0,
NN\,202\,48\,4339 and NN\,202\,23\,1837), the Polish National Center
for Science (Grants~2011\slash03\slash N\slash ST2\slash03691, 
2013\slash11\slash N\slash ST2\slash03879, 
2014\slash13\slash N\slash ST2\slash02565,
2014\slash14\slash E\slash ST2\slash00018
and
2015\slash18\slash M\slash ST2\slash00125), 
the Foundation for Polish Science --- MPD program, co-financed by the
European Union within the European Regional Development Fund, the
Federal Agency of Education of the Ministry of Education and Science
of the Russian Federation (SPbSU research Grant 11.38.242.2015), the
Russian Academy of Science and the Russian Foundation for Basic
Research (Grants 08-02-00018, 09-02-00664 and 12-02-91503-CERN), the
Ministry of Education, Culture, Sports, Science and Tech\-no\-lo\-gy,
Japan, Grant-in-Aid for Sci\-en\-ti\-fic Research (Grants 18071005,
19034011, 19740162, 20740160 and 20039012), the German Research
Foundation (Grant GA\,1480/2-2), the EU-funded Marie Curie Outgoing
Fellowship, Grant PIOF-GA-2013-624803, the Bulgarian Nuclear
Regulatory Agency and the Joint Institute for Nuclear Research, Dubna
(bilateral contract No. 4418-1-15\slash 17), Ministry of Education and
Science of the Republic of Serbia (Grant OI171002), Swiss
Nationalfonds Foundation (Grant 200020\-117913/1), ETH Research
Grant TH-01\,07-3 and the U.S.\ Department of Energy.

\bibliographystyle{na61Utphys}
\bibliography{main.bbl}

\newpage
{\Large The \NASixtyOne Collaboration}
\bigskip
\begin{sloppypar}
  
\noindent
A.~Aduszkiewicz$^{\,17}$,
Y.~Ali$^{\,15,32}$,
E.~Andronov$^{\,23}$,
T.~Anti\'ci\'c$^{\,3}$,
N.~Antoniou$^{\,8}$,
B.~Baatar$^{\,21}$,
F.~Bay$^{\,25}$,
A.~Blondel$^{\,27}$,
M.~Bogomilov$^{\,2}$,
A.~Bravar$^{\,27}$
% ...

\end{sloppypar}
\noindent
% ...
$^{2}$~Faculty of Physics, University of Sofia, Sofia, Bulgaria\\
$^{3}$~Ru{\dj}er Bo\v{s}kovi\'c Institute, Zagreb, Croatia\\
% ...
$^{8}$~University of Athens, Athens, Greece\\
% ...
$^{15}$~Jagiellonian University, Cracow, Poland\\
% ...
$^{17}$~University of Warsaw, Warsaw, Poland\\
% ...
$^{21}$~Joint Institute for Nuclear Research, Dubna, Russia\\
% ...
$^{23}$~St. Petersburg State University, St. Petersburg, Russia\\
% ...
$^{25}$~ETH Z\"urich, Z\"urich, Switzerland\\
% ...
$^{27}$~University of Geneva, Geneva, Switzerland\\
% ...
$^{32}$~Present address: COMSATS Institute of Information Technology, Islamabad, Pakistan\\
% ...

\end{document}